\documentclass[12pt,english]{qf}
\usepackage[T1]{fontenc}
\usepackage[latin9]{inputenc}
\usepackage{array}
\usepackage{booktabs}
\usepackage{multirow}
\usepackage{amsmath}
\usepackage{amssymb}
\usepackage{graphicx}
\usepackage[authoryear]{natbib}
\usepackage{xargs}[2008/03/08]

\makeatletter

\providecommand{\tabularnewline}{\\}

\usepackage[nomarkers,tablesfirst,nolists]{endfloat}

\makeatother

\usepackage{babel}
\begin{document}

\title{Decision trees unearth return sign correlation in the S\&P 500}

\author{L. Fi\'evet$*\dagger$\thanks{$^\ast$Corresponding author. Email: lfievet@ethz.ch}
and D. Sornette$\dagger\ddagger$\affil{
$\dag$Chair of Entrepreneurial Risks, Department of Management, Technology, and Economics, ETH, Scheuchzerstrasse 7, 8092 Z\"urich, Switzerland\\
$\ddag$Swiss Finance Institute, Z\"urich, Switzerland}
\received{Received 14 April 2017}}
\maketitle
\begin{abstract}
\noindent {\footnotesize{}Technical trading rules and linear regressive
models are often used by practitioners to find trends in financial
data. However, these models are unsuited to find non-linearly separable
patterns. We propose a decision tree forecasting model that has the
flexibility to capture arbitrary patterns. To illustrate, we construct
a binary Markov process with a deterministic component that cannot
be predicted with an autoregressive process. A simulation study confirms
the robustness of the trees and limitation of the autoregressive model.
Finally, adjusting for multiple testing, we show that some tree based
strategies achieve trading performance significant at the 99\% confidence
level on the S\&P 500 over the past 20 years. The best strategy breaks
even with the buy-and-hold strategy at 21 bps in transaction costs
per round trip. A four-factor regression analysis shows significant
intercept and correlation with the market. The return anomalies are
strongest during the bursts of the dotcom bubble, financial crisis,
and European debt crisis. The correlation of the return signs during
these periods confirms the theoretical model.}{\footnotesize \par}
\end{abstract}
\begin{classcode}
C01, C12, C14, C15, C22, G01, G14, G17
\end{classcode}

\pagebreak{}

\global\long\def\lag{\varrho}

\section{Introduction}

The efficient market hypothesis \citep{malkiel_fama_1970,fama_1991}
is a cornerstone of theoretical finance that has been widely debated
and tested. The hypothesis states that market prices instantaneously
and fully reflect all available information. Mathematically speaking,
market returns follow a martingale process after adjusting for equilibrium
expected returns. As a consequence, in an efficient market it is impossible
to develop trading strategies that are consistently profitable in
excess of the buy-and-hold strategy. However, real markets do not
meet some necessary conditions for the Efficient Market Hypothesis
(EMH) such as no transaction costs, and cost-less availability of
information to all market participants. In contrast, real markets
are better described as being in an equilibrium state of disequilibrium
\citep{grossmanstiglitz1980}, in which a perpetual flow of new events
needs to be constantly arbitraged to keep markets marginally efficient.
In particular, the recurring occurrence of bubbles and crashes (i.e.
dotcom bubble, financial crisis, Chinese stock market turbulence)
supports the hypothesis that irrational investors can provoke large
departures from rational and efficient markets \citep{Johansen2000a,emhcritics,sornettebook03,harrassornette2011}.

The EMH cannot be proven true in general, but can be proven false
by finding a sensible trading strategy significantly profitable in
excess of the benchmark. To reject the null hypothesis of efficient
markets, \citet{White2000} proposed the reality check method that
computes the statistical significance of the best trading strategy
in a universe of strategies. The reality check method determines if
the best strategy outperforms a given benchmark strategy, while accounting
for the exact dependency structure of all tested strategies. This
method has since been extended by \citet{romanowolf05} to reject
as many null hypotheses as possible while controlling for the familywise
error rate. As well, they advocate the use of studentized test statistics
to improve robustness in finite samples. Refinements for Heteroskedasticity
and Auto-Correlation (HAC) robust performance metrics have been established
by \citet{RePEc:zur:iewwpx:320}.

The existing studies assessing the EMH using a multiple-testing framework
provide a heterogeneous picture. The study by \citet{White2000} finds
no significant technical trading rule during a three year time span
on the S\&P 500. This result is confirmed by \citet{Sullivan1999}
on a 10 year time period on the S\&P 500, however they find significant
trading rules on the Dow Jones for a 100 year time period ending in
1996. Subsequently, \citet{citeulike:322540} analyze an extended
universe of strategies and find no significantly performing strategy
on the Dow Jones and S\&P 500 during the period 1990-2000. However,
they find significant trading rules for the more recent NASDAQ and
Russell 2000 during the same period. Later, \citet{Hsu2010} compare
the pre-ETF and post-ETF period for the U.S. and emerging markets.
The post-ETF period is found to be fairly efficient, and in both periods
emerging markets are found to be less efficient then the more mature
U.S. markets. The efficiency check of foreign exchange markets by
\citet{Hsu2016} reveals that technical trading has been profitable
in the past, but returns have been declining. As for stock markets,
the foreign exchanges of emerging countries are less efficient then
in developed countries.

The picture is equally mixed when analyzing funds. The analysis by
\citet{fama_french_2009} shows that during the period of 1984 to
2006 mutual funds underperform in aggregate the three-factor, and
four-factor benchmarks by about the transaction costs. When looking
at the individual performance, even the best performers are not statistically
significant. The study of \citet{Barras16} mitigates this picture,
finding significantly performing funds prior to 1996, and almost none
afterwards. This result supports the stock market studies discussed
above that showed increasing efficiency over time. In contrast, the
portfolio approach by \citet{Yen2015} to measure fund performance
does find significant performance during a 10 year period ending in
2007.

We note that the finding of statistically significant strategies does
not necessarily reject the EMH. A statistically significant strategy
is a necessary condition to reject the EMH, but it is not sufficient
as traders may have been unable to select the best strategy ex-ante.
To cover this aspect, the EMH definition of \citet{Timmermann200415}
requires a set of search technologies among which one would have selected
the strategy to use. In practical terms, there must exist a sensible
search technology that would have selected a profitable strategy.
A search technology can as well be a portfolio of strategies that
is rebalanced periodically, and generates statistically significant
profits.

The existing assessments of the EMH are focused on testing the performance
of technical trading strategies and funds. However, the anomalies
reported by \citet{Niederhoffer1966}, \citet{Zhang1999}, \citet{Leung2000},
\citet{andersensornettepocket05}, \citet{Zunino2009}, \citet{satinover_sornette_2012,satinover_sornette_2012_2}
and \citet{isl} have not undergone a rigorous analysis to the best
of our knowledge. These anomalies are particularly interesting because
many linearly inseparable patterns, as for example the logical exclusive-OR
function (XOR), would remain undetected with linear models such as
moving averages. Nonetheless, such patterns can be detected using
statistical learning models such as decision trees.

In this paper, we demonstrate the limitation of autoregressive models
to capture the predictable XOR pattern, and show how decision tree
based models overcome this limitation. We carefully discuss different
variants of decision trees and how to avoid the issue of overfitting.
Especially, we derive the connection between fixed decision trees
and Markov chains. The connection is used to derive a binary Markov
process of order $\lag$, analogous to the autoregressive process
of order $\lag$\footnote{We use $\lag$ as order parameter (i.e. lag) instead of the common
$p$ to avoid confusions with probabilities denoted by $p$.}, but with more intricate non-linear autocorrelation patterns. We
then prove that the parameters of an autoregressive process of order
$\lag$ have zero expectation for a binary Markov process satisfying
certain conditions.

To confirm the theoretical results, we simulate the statistical significance
of the competing models on an autoregressive process of order two
and a binary Markov process of order two. The statistical significance
is studied based on the studentized Sharpe ratio as a function of
the sample size, the calibration window length, and the autocorrelation
parameters. The simulation results show that the autoregressive forecast
is optimal on the autoregressive process, but has no forecasting power
on the binary Markov process. The fixed regression and classification
trees are more robust, performing equally well on both processes.
The regression and classification trees with dynamically optimized
decision boundaries are prone to overfitting and underperform the
fixed trees.

Finally, all competing model are tested on daily returns of the S\&P
500 from Jan. 1, 1995 to Dec. 31, 2015. The multiple testing adjusted
and HAC robust statistical significance of a trading strategy, derived
from a forecasting model, is computed by combining the methodologies
described in \citet{romanowolf05} and \citet{RePEc:zur:iewwpx:320}.
Each model is tested for a range of lags and calibration window length,
leading to a universe of 1000 strategies. The null hypothesis of no
predictability for the best strategy is rejected at the 99\% confidence
level. The best performing model breaks even with the buy-and-hold
strategy at transaction costs per round trip of 21.4 bps. This profitability
threshold exceeds the 5 bps in transaction costs applied nowadays
on the future market.

This paper continues as follows, Section \ref{sec:methods} presents
the competing autoregressive and tree based models. Section \ref{sec:Comparing-Forecasting-Power}
provides analytical comparisons for the forecasting power of the autoregressive
and tree based models. Section \ref{sec:Statistical-Test} presents
the HAC robust multiple testing methodology used to computed p-values
adjusted for data snooping. Sections \ref{sec:Simulation-Study} and
\ref{sec:Empirical-Results} show the simulation results, respectively
the empirical results. The last section concludes and outlines a path
for future research.

\section{Motivating Tree Based Forecasting\label{sec:methods}}

\newcommandx\AR[1][usedefault, addprefix=\global, 1=\lag]{\textrm{AR}(#1)}
\newcommandx\ARMA[2][usedefault, addprefix=\global, 1=\lag, 2=q]{\textrm{ARMA}(#1,\,#2)}
\newcommandx\MA[1][usedefault, addprefix=\global, 1=\lag]{\textrm{MA}(#1)}
\newcommandx\BMP[1][usedefault, addprefix=\global, 1=\lag]{\textrm{M}(#1)}
\newcommandx\arparam[1][usedefault, addprefix=\global, 1=]{\phi_{#1}}
\newcommandx\innovation[1][usedefault, addprefix=\global, 1=t]{a_{#1}}
\global\long\def\tree{\mathcal{T}}
\global\long\def\bincat{\left\{  -,\,+\right\}  }
\newcommandx\prob[1][usedefault, addprefix=\global, 1=i]{p_{#1}}
\newcommandx\outplus[1][usedefault, addprefix=\global, 1=i]{\prob[#1+]}
\newcommandx\outminus[1][usedefault, addprefix=\global, 1=i]{\prob[#1-]}
\newcommandx\inplus[1][usedefault, addprefix=\global, 1=i]{\prob[+#1]}
\newcommandx\inminus[1][usedefault, addprefix=\global, 1=i]{\prob[-#1]}
\newcommandx\Deltaprob[1][usedefault, addprefix=\global, 1=i]{\Delta\prob[#1]}
\newcommandx\stationary[1][usedefault, addprefix=\global, 1=]{\pi_{#1}}
\newcommandx\stochvar[1][usedefault, addprefix=\global, 1=]{X_{#1}}
\newcommandx\stochprocess[1][usedefault, addprefix=\global, 1=t]{\left\{  \stochvar[#1]\right\}  }
\global\long\def\samples{\mathbf{X}}
\global\long\def\slinputs{\boldsymbol{X}}
\newcommandx\slinput[1][usedefault, addprefix=\global, 1=t]{x_{#1}}
\global\long\def\sloutputs{\boldsymbol{Y}}
\newcommandx\sloutput[1][usedefault, addprefix=\global, 1=t]{y_{#1}}
\newcommandx\state[2][usedefault, addprefix=\global, 1=i, 2=]{S_{#1}^{#2}}
\global\long\def\states{\boldsymbol{S}}
\global\long\def\time{T}
\global\long\def\strategies{\mathcal{S}}
\newcommandx\region[1][usedefault, addprefix=\global, 1=]{R_{#1}}
\newcommandx\regions[1][usedefault, addprefix=\global, 1=n]{\left\{  \region[1],\,\ldots,\,\region[#1]\right\}  }
\newcommandx\return[1][usedefault, addprefix=\global, 1=t]{r_{#1}}
\newcommandx\regionmean[1][usedefault, addprefix=\global, 1=i]{c_{#1}}
\global\long\def\bandalpha{a}
\newcommandx\bandwidth[2][usedefault, addprefix=\global, 1=T, 2=]{S_{#1}^{#2}}
\global\long\def\pvalue{\rho}
\global\long\def\model{\mathcal{M}}
\global\long\def\signal{\mathfrak{s}}
\global\long\def\calibrationL{L}
\global\long\def\betamarket{\beta_{MKT}}
\global\long\def\betahml{\beta_{HML}}
\global\long\def\betasmb{\beta_{SMB}}
\global\long\def\betamom{\beta_{MOM}}
\global\long\def\intercept{\alpha}
\newcommandx\riskfree[1][usedefault, addprefix=\global, 1=]{r_{#1}^{f}}
\newcommandx\rmarket[1][usedefault, addprefix=\global, 1=]{r_{#1}^{MKT}}

\subsection{Linear Filter Models \& Technical Trading Rules}

Wold's decomposition theorem \citep{Mills1999,hamilton_1994} states
that every weakly stationary, purely non-deterministic stochastic
process $\stochprocess$ can be written as a linear filter with infinite
lag. All time series analysis models defined as a finite-order stochastic
difference equation derive from this general concept. The simplest
stationary models being the autoregressive model $\AR$ and moving
average model $\MA$ of order $\lag$. Non-stationary models can be
built using nested stationary models for the mean and variance.

These models of stochastic processes are foremost characterized by
the autocorrelation function of returns and absolute returns. While
these characteristics are predominantly used to described financial
returns, several stylized facts such as gain/loss asymmetry and heavy
tails are well document \citep{cont_2001}. The gain/loss asymmetry
is often described by the skewness of the return distribution and
the heavy tails by the kurtosis. However, possible non-linear dependencies
are often neglected.

Non-linear stochastic processes arise when their representation is
obtained by some non-linear mechanism, for example polynomial dependencies
in the innovation, asymmetric innovations, correlated innovations,
time varying parameters, or regime switching. Unfortunately, testing
for non-linearity is a challenging task and not possible in general
when the functional form of the non-linearity is unknown \citep[chap. 6]{Mills1999}.
Past efforts have been concentrated on modeling the common stylized
facts such as skewness, fat tails, volatility clustering, and regime
switching, using non-linear combinations of linear filter models \citep{MillsEncyclopedia}.
This continued use of linear models as building blocks for non-linear
models leaves the possibility that deterministic non-linearly separable
patterns have remained undetected in past studies

To forecast trends in stock markets, traders have developed an extensive
taxonomy of technical trading rules expressing their beliefs about
the market behavior. Some of these rules rely on a linear filter model
such as moving average, which forecasts a trend persistency. While
a majority of rules define trading triggers based on support and resistance
bands, channels, and oscillators, these rules do not focus on a potential
intrinsic structure of market returns, but on prices levels psychologically
important to a large number of traders. Assuming that a majority of
market participants acts based on these price levels, these rules
should have significant profitability.

A common feature of linear filter models and technical trading rules
is that they do not test for some potential non-linear dependencies.
An example are the return sign correlation, which have been well documented
by \citet{Christoffersen2003} and \citet{Christoffersen}. In this
section, we argue that return sign correlations can remain undetected
in the autocorrelation function, and propose decision trees as a forecasting
model to detect return sign correlations.

\subsection{From Autoregressive to Tree Based Models}

\begin{figure}[p]
\begin{centering}
\includegraphics[width=0.5\textwidth]{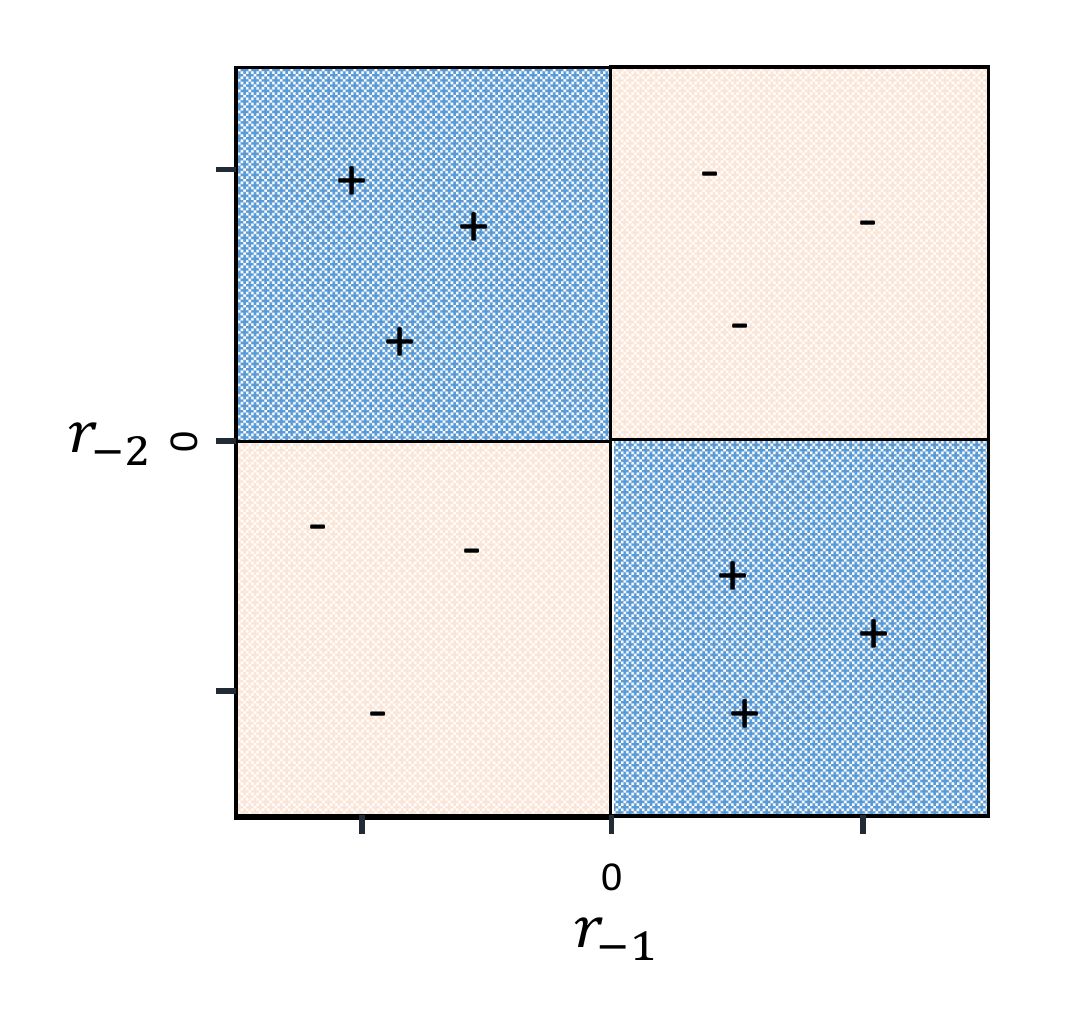}
\par\end{centering}
\caption{\textbf{Illustration of the XOR pattern\label{fig:xor-pattern}}}
\end{figure}

The autoregressive model $\AR$ defines the evolution of the time
series $\stochprocess$ with $\lag$ lags as

\begin{equation}
\stochvar[t]=\arparam[0]+\sum_{i=1}^{\lag}\arparam[i]\stochvar[t-i]+\innovation,\label{eq:arp}
\end{equation}
with parameters $\arparam=\left(\arparam[0],\,\arparam[1],\,\ldots,\,\arparam[\lag]\right)$,
and i.i.d. innovations $\innovation$. The shortcoming of autoregressive
models can be illustrated with the two argument exclusive-OR function
$XOR(\return[-1],\,\return[-2])$ that returns true ($=1$) when exactly
one of the arguments is true and false ($=-1$) otherwise (see Figure
\ref{fig:xor-pattern}). An example of XOR like data is
\begin{equation}
\samples=\left\{ \left(\left(1,\,1\right),\,-1\right),\,\left(\left(1,\,-1\right),\,1\right),\,\left(\left(-1,\,1\right),\,1\right),\,\left(\left(-1,\,-1\right),\,-1\right)\right\} ,\label{eq:xor-data}
\end{equation}
where the four samples are assumed to be at independent times. The
notation of Equation (\ref{eq:xor-data}) is taken from the statistical
learning literature \citep{isl}, where $\samples$ is the training
data available, and $\left(\left(\stochvar[t-2],\,\stochvar[t-1]\right),\,\stochvar[t]\right)$
denotes the sample at time $t$ with input $\slinput=\left(\stochvar[t-2],\,\stochvar[t-1]\right)$
and output (=response) $\sloutput=\stochvar[t]$. Calibrating the
autoregressive model of order two to the data of Equation (\ref{eq:xor-data})
yields the parameters
\begin{equation}
\left(\arparam[0]=0,\,\arparam[1]=0,\,\arparam[2]=0\right),
\end{equation}
which fail at capturing the deterministic XOR function. As we will
discuss in section \ref{subsec:Binary-Markov-DGP}, almost XOR like
patterns can arise in time series data.

Non-linearly separable patterns can be modeled using a partition $\region=\regions[n]$
of the input (or feature) space into $n$ regions, and assigning the
constant values $\left\{ \regionmean[1],\,\ldots,\,\regionmean[n]\right\} $
to each region. The resulting evolution of the time series $\stochvar[t]$
can then be written as

\begin{equation}
\sloutput=\stochvar[t]=\sum_{i=1}^{n}\regionmean\cdot I\left\{ \slinput\in\region[i]\right\} +\innovation,\label{eq:regressive-region-process}
\end{equation}
where $I$ is the indicator function, and $\innovation$ are i.i.d.
innovations. This modeling approach allows for an arbitrary flexibility,
as any function can be approximated to any precision with a sufficient
number of regions. For example, the XOR data from Equation (\ref{eq:xor-data})
can be modeled exactly by the two regions $\region[1]=\left\{ x\in\mathbb{R}^{2}|x_{1}x_{2}\geq0\right\} $
with $\regionmean[1]=-1$, and $\region[2]=\left\{ x\in\mathbb{R}^{2}|x_{1}x_{2}<0\right\} $
with $\regionmean[2]=1$. The downside of this modeling approach is
that the number of parameters increases arbitrarily with the number
of regions, and a procedure to control for overfitting is required.

What remains unspecified in the stochastic process of Equation (\ref{eq:regressive-region-process})
is the algorithm to estimate the regions based on a given realization
$\stochprocess_{1}^{\time}$ of a process. In general, regions of
arbitrary shape and overlaps can be used. However, for the purpose
of this study, decision tree models provide sufficient flexibility.
Decision trees find rectangular regions, using a recursive splitting
algorithm of the input space that minimizes a loss function for the
given training data. While several variations of the splitting algorithms
exist, we will focus on the most popular Classification And Regression
Tree (CART) algorithm described by \citet[chap. 9]{esl}.

In the context of financial returns with low signal to noise ratio,
finding the best tree is an NP-complete problem \citep{HYAFIL197615}
that is not computationally feasible. This stands in stark contrast
to autoregressive models were the global optimum can be estimated.
However, this is not an issue as the CART algorithm is nonetheless
deterministic by using recursive binary splitting, and subsequent
pruning. The algorithm always finds the same tree for a given training
data, allowing practitioners to independently obtain the same forecast.
The definition of the splitting algorithms for regression and classification
are given below, as well as a discussion of robustness in the context
of highly stochastic data.

\subsection{Regression Tree Algorithm\label{subsec:Regression-Tree}}

The CART algorithm constructs a regression tree using the Mean Squared
Error (MSE) as the loss function $Q_{MSE}\left(y,\,\hat{y}\right)=\left(y-\hat{y}\right)^{2}$.
For a given training data $\samples=\left\{ \left(x_{1},\,y_{1}\right),\,\ldots,\,\left(x_{N},\,y_{N}\right)\right\} $
and partitioning of the input space into regions $\region=\left\{ \region[1],\,\ldots,\,\region[n]\right\} $,
the algorithm assigns the response
\begin{equation}
\hat{c}_{i}=\left\langle \left\{ \sloutput[i]|\slinput[i]\in\region[i]\right\} \right\rangle 
\end{equation}
to region $\region[i]$, which is the mean observed output in that
region. The binary splitting algorithm adds one new region at each
iteration by splitting an existing region into two. At each iteration
an existing region $\region$ is split into two halves with a plane
defined by variable $j$ and split point $s$ as
\begin{equation}
R^{1}\left(j,\,s\right)=\left\{ x|x_{j}\leq s\right\} \:\textrm{and}\:R^{2}\left(j,\,s\right)=\left\{ x|x_{j}>s\right\} .
\end{equation}
 The optimal split is given by
\begin{equation}
\min_{j,\,s}\left[\sum_{x_{i}\in R^{1}}Q\left(y_{i},\,\hat{c}_{1}\right)+\sum_{x_{i}\in R^{2}}Q\left(y_{i},\,\hat{c}_{2}\right)\right],
\end{equation}
which can be determined efficiently by running through all the input
variables and split points defined by the inputs.

In many scenarios, for example of XOR like training data, a split
with not improvement in the loss can be followed by a split with a
large reduction of the loss. This prevents the introduction of termination
criterion in the splitting procedure. To overcome this issue, the
fully grown tree $\tree_{0}$ is subsequently pruned. The pruning
procedure finds an optimal subtree $\tree$ without the internal nodes
that do not affect significantly the loss. The loss of a subtree $\tree\subset\tree_{0}$
is given by
\begin{equation}
Q\left(\tree,\,\alpha\right)=\sum_{m=1}^{\left|\tree\right|}N_{m}Q_{m}\left(t\right)+\alpha\left|\tree\right|,\label{eq:tree-pruning}
\end{equation}
where $Q_{m}\left(\tree\right)=\frac{1}{N_{m}}\sum_{x_{i}\in R_{m}}Q\left(y_{i},\,\hat{c}_{m}\right)$
is the loss of the terminal node $m$, $N_{m}=\#\left\{ x_{i}\in R_{m}\right\} $
is the number of samples in that node, and $\left|\tree\right|$ is
the number of terminal nodes in the tree. The regularization coefficient
$\alpha$ of the tree size determines to which degree smaller trees
are favored with respect to higher loss. The value $\alpha=0$ would
yield the fully grown tree $\tree_{0}$.

In the context of predicting financial returns, the signal to noise
ratio in the samples is intrinsically low, and without appropriate
pruning the data would be overfit. However, the Scikit-learn library
\citep{scikit-learn} used in this paper does not use a pruning parameter
to control for overfitting, but instead allows us to set a lower bound
on the number of samples in a terminal node. Fortunately, determining
the adequate number of samples in a terminal node is simpler then
determining an equivalent pruning parameter. The lower bound on the
number of samples in a terminal node of a classification tree is computed
in the following Section \ref{subsec:Classification-Tree}. The lower
bound for a regression tree is set to the lower bound value of the
equivalent classification tree.

We remark that the objective of the MSE loss function, namely minimizing
variance within each region, is not necessarily optimal for forecasting
financial returns. The suboptimality arises when the returns within
a region exhibit low variance and mean close to zero. The insignificant
mean implies that no forecast can be made. However, this does not
negatively impact performance as return forecasts of small amplitude
can be discarded as insignificant.

\subsection{Classification Tree\label{subsec:Classification-Tree}}

The dominantly stochastic behavior of financial returns translates
into particularly low statistical significance of regression model
forecasts. However, the statistical significance of a potential signal
can eventually be improved by mapping the outputs to categorical outputs.
Under the assumption that the mapping removes more noise then signal,
the resulting classification problem is more robust. The simplest
such mapping is $r\rightarrow\textrm{sign}\left(r\right)$ that results
in predicting only an up and down moves. The CART algorithm described
in Section \ref{subsec:Regression-Tree} only needs one modification
for classification, namely a different loss function because the MSE
loss is not suitable for classification. Typically, the classification
of $K$ classes (e.g. $\bincat$ with $K=2$) is performed using the
Gini index $Q_{Gini}\left(\vec{p}\right)=\sum_{i\neq j}^{K}\prob\prob[j]$,
which maximizes the probability $\prob$ of a single class in a region.

In the context of predicting market up or down moves, the Gini index
naturally sets the right objective, as it will maximize the predictability
of an up or down move in a given region. However, the Gini index by
itself is prone to overfitting, as terminal nodes with a few samples
of the same kind are favored despite such nodes being a likely occurrence
under the null hypothesis of no predictability. To alleviate this
problem, we set a lower bound on the number of samples in a terminal
node.

The probability distribution of $k$ up moves ($\uparrow$) and $n-k$
down moves ($\downarrow$), assuming equal class probability, is given
by the binomial distribution
\begin{equation}
P\left(\#\uparrow=k,\,\#\downarrow=n-k\right)=\left(\frac{1}{2}\right)^{n}\left(\begin{array}{c}
n\\
k
\end{array}\right).
\end{equation}
For sufficiently large $n$, this binomial distribution can be approximated
by the normal distribution $\mathcal{N}\left(\frac{n}{2},\,\frac{1}{4n}\right)$.
Now we compute $n$ assuming that the prediction is significant at
the one standard deviation and the class probabilities satisfy $\left|\prob[+]-\prob[-]\right|=2\cdot0.05$.
This implies $\frac{1}{4n}=0.05^{2}$, or $n\geq100$, which would
require at least 100 samples per terminal node. This illustrates the
difficulty of obtaining statistically significant forecasts and the
large amount of calibration data required. In general, the minimum
number of samples per terminal node should be picked as large as possible.

\subsection{Fixed Tree\label{subsec:Fixed-Tree}}

Similarly to the classification presented in Section \ref{subsec:Classification-Tree},
the signal to noise ratio in the input variables can be improved by
a mapping of the inputs onto the two categories $\left\{ -,\,+\right\} $.
In case of inputs with $\lag$ lags, this results in the discrete
input space $\left\{ -,\,+\right\} ^{\lag}$ with $2^{\lag}$ elements.
For a training set $\mathcal{D}$, the limited number of possible
inputs controls the expected number of samples in a leaf $\region$
as $E\left[N_{R}\right]=\frac{\left|\mathcal{D}\right|}{2^{p}}$.
Therefore, with an appropriate choice of $\lag$ with respect to the
sample size $\left|\mathcal{D}\right|$, the problem of overfitting
does not arise. This allows us to remove the lower bound on the number
of samples per leaf, resulting in the CART algorithm to generate the
fully grown tree with regions $\regions[2^{\lag}]$ (i.e. one region
for each input). In other terms, the tree is fixed by the number of
lags $\lag$, and the CART algorithm reduces to compute the class
probabilities in each leaf. A fixed tree can be used for classification,
prediction of the majority vote inside each leaf, or regression by
predicting the mean return of a leaf.

\subsection{Probit Models \& Polynomial Features}

Probit models could be used as the autoregressive equivalent of classification
trees, calibrating the parameters using only binary up and down returns.
Nonetheless, just as the autoregressive model, the probit model cannot
capture linearly inseparable patterns like the XOR function. To capture
non-linear patterns, the autoregressive or probit models have to be
extended with non-linear features. For example, the XOR function can
be described with the quadratic feature $\return[t-1]\cdot\return[t-2]$. 

However, polynomial features are not robust with respect to skewed
distributions or multiple patterns offset with respect to the origin
(e.g. offset as $\left(\return[t-1]-x_{1}\right)\left(\return[t-2]-x_{2}\right)$).
In contrast, decision trees are non-linear predictors robust with
respect to both issues. The CART algorithm can find localized non-linear
predictability independently of its location in the input space, and
use multiple regions to approximate skewness.

\section{Comparing Forecasting Power\label{sec:Comparing-Forecasting-Power}}

\subsection{Connecting Fixed Trees to Markov Chains\label{subsec:Connecting-Fixed-Trees}}

Given a fixed decision tree, the time evolution defined by Equation
(\ref{eq:regressive-region-process}) defines as well the transition
probability $p$ from the sample at time $t$ to the sample at time
$t+1$ as
\begin{equation}
\left(\left(\stochvar[t-\lag],\,\ldots,\,\stochvar[t-1]\right),\,\stochvar[t]\right)\overset{p}{\rightarrow}\left(\left(\stochvar[t-\lag+1],\,\ldots,\,\stochvar[t]\right),\,\stochvar[t+1]\right).
\end{equation}
The sample inputs $\left(\stochvar[t-\lag],\,\ldots,\,\stochvar[t-1]\right)$
define $2^{\lag}$ possible states $\left\{ \state[1],\,\ldots,\,\state[2^{\lag}]\right\} $,
and the time evolution defines the two possible transitions from each
state to another state, while the remaining $2^{\lag}-2$ states are
unattainable. In other terms, the fixed decision tree defines a $\lag$-th
order Markov chain with $2^{\lag}$ states, and two time dependent
non-zero outgoing transition probabilities from each state.

We denote by $\boldsymbol{P}\in\left[0,\,1\right]^{2^{\lag}\times2^{\lag}}$
the time independent transition matrix of the Markov chain defined
by a fixed tree $\tree$, resulting from the mapping of the returns
onto the states $\bincat$. The elements $\boldsymbol{P}_{ij}$ of
the transition matrix describe the probabilities $P\left(\state[j]\rightarrow\state[i]\right)$
to go from state $\state[j]$ to state $\state[i]$. The probabilities
to transition from one state to any of the states must always sum
to one, imposing $\sum_{i=1}^{2^{\lag}}\boldsymbol{P}_{ij}=1,\,\forall j$.
Due to the particular structure of the tree, each state has two incoming
transitions, and two outgoing transitions. This implies that each
row and column in $\boldsymbol{P}$ has two non-zeros entries.

This Markov chain admits a stationary distribution if there exists
a vector of probabilities $\stationary\in\left[0,\,1\right]^{2^{\lag}}$,
satisfying $\left|\stationary\right|=\sum_{i=1}^{2^{\lag}}\left|\stationary[i]\right|=1$
and
\begin{equation}
\boldsymbol{P}\stationary=\stationary.\label{eq:markov-stationary}
\end{equation}
The component $\stationary[i]$ of the vector $\stationary$ represents
the probability to be in state $\state[i]$ at the stationary regime.
The equality in Equation (\ref{eq:markov-stationary}) asserts the
stationarity condition that the probability of each state is invariant
when transitioning to the next state. The stationary distribution
$\stationary$ is an eigenvector of $\boldsymbol{P}$ with unit eigenvalue.
The eigenvalues $\left\{ \lambda_{1},\,\ldots,\,\lambda_{2^{\lag}}\right\} $
of $\boldsymbol{P}$ all satisfy $0\leq\lambda_{i}\leq1$, and consequently
the stationary distribution is always determined by the eigenstates
(i.e. eigenvectors) with unit eigenvalues, while the eigenstates associated
to eigenvalues $\lambda<1$ decay away. The second largest eigenvalue
determines how quickly the stationary distribution is reached. Multiple
eigenstates with unit eigenvalue arise in the case of reducible chains
composed of several independent Markov chains. This can be shown by
the orthogonality property of eigenstates and all their components
being strictly positive in the context of Markov chains.

\subsection{Binary Markov Based Processes\label{subsec:Binary-Markov-DGP}}

In the case of binary categories $\bincat$, namely up ($\stochvar[t]\geq0$)
and down ($\stochvar[t]<0$) moves, the Markov chain defined in Section
\ref{subsec:Connecting-Fixed-Trees} has $2^{\lag}$ possible states.
Each state $\state[i]\in\bincat^{\lag}$ has two outgoing transitions
probabilities $\outplus$ and $\outminus$, the probability of an
up move, respectively a down move. These two outgoing probabilities
are subject to $\outplus+\outminus=1$, and we can characterize them
by a single variable $\Deltaprob=\outplus-\outminus$, where $\outplus=\frac{1}{2}\left(1+\Deltaprob\right)$
and $\outminus=\frac{1}{2}\left(1-\Deltaprob\right)$. Each state
$S_{i}$ has as well two incoming transition probabilities, which
we denote by $\inplus=P\left(\state[+i]\rightarrow\state[i]\right)$
and $\inminus=P\left(\state[-i]\rightarrow\state[i]\right)$, where
the first index ($+$ or $-$) stands for the first sign of the previous
state (e.g. $S_{i}=\left(-,\,+\right)$, $S_{+i}=\left(+,\,-\right)\rightarrow S_{i}$,
and $S_{-i}=\left(-,\,-\right)\rightarrow S_{i}$). We remark that
the sets of outgoing and incoming probabilities are in one-to-one
correspondence. The duplicate definition of the transition probabilities
is introduced for subsequent convenience.

The stationary distribution $\stationary$ satisfies
\begin{equation}
\stationary[+i]\cdot\inplus+\stationary[-i]\cdot\inminus=\stationary[i],\,\forall1\leq i\leq2^{\lag},
\end{equation}
where $\stationary[+i]$ and $\stationary[-i]$ stand for the probabilities
at stationarity of the state $S_{+i}$, respectively $\state[-i]$.
We remark that the outgoing probabilities of each state must always
sum to one, but the incoming probabilities can be arbitrary $\inplus,\,\inminus\in\left[0,\,1\right]$.
In the special case of incoming probabilities $\inplus+\inminus=1$,
the stationary distribution is given by $\stationary[i]=\frac{1}{2^{\lag}},\,\forall i$.
In other terms, the stationary distribution with uniform probability
for all states arises when the incoming probabilities of each state
(i.e. rows in $\boldsymbol{P}$) sum to one. This result can be derived
from Equation (\ref{eq:markov-stationary}) by setting $\stationary=\frac{1}{2^{\lag}}\vec{1}$,
which leads to $\boldsymbol{P}\frac{1}{2^{\lag}}\vec{1}=\frac{1}{2^{\lag}}\vec{1}\Rightarrow\sum_{j=1}^{2^{\lag}}\boldsymbol{P}_{ij}=\inplus+\inminus=1,\,\forall i$.

Based on the Markov chain, we define the Data Generating Process (DGP)
\begin{equation}
X_{t+1}=\begin{cases}
+\left|\mathcal{N}\left(\innovation[t+1]|\,0,\,\sigma\right)\right| & \textrm{with probability }\outplus[t]\\
-\left|\mathcal{N}\left(\innovation[t+1]|\,0,\,\sigma\right)\right| & \textrm{with probability }\outminus[t]
\end{cases},\label{eq:Markov-DGP}
\end{equation}
where $\outplus[t]$ and $\outminus[t]$ are the outgoing transition
probabilities for the state $\state[t]$ at time $t$.

In order for this DGP to generate a continuous normal distribution,
without jump at $\stochvar[t]=0$, the number of up and down moves
needs to be equal, which is enforced by the condition
\begin{equation}
B\left(\boldsymbol{P}\right)=\sum_{i=1}^{2^{\lag}}\stationary[i]\Deltaprob=0.\label{eq:balancing-equation}
\end{equation}
As the stationary vector $\stationary$ is a function of $\boldsymbol{P}$,
the Equation (\ref{eq:balancing-equation}) is a non trivial polynomial
equation in $\boldsymbol{P}$. However, in the case where all states
are equally likely, the balance of up and down moves is achieved when
$\sum_{i=1}^{2^{\lag}}\Deltaprob=0$.

We remark that the balancing condition is not the only possibility
to obtain a continuous distribution. The balance between up and down
moves can be broken locally as long as the balancing Equation (\ref{eq:balancing-equation})
has expectation zero: $E\left[B\left(\boldsymbol{P}\right)\right]=0$.
Assuming sufficiently low autocorrelation over time for the balancing
term $B\left(\boldsymbol{P}\right)$, the discontinuity of the return
distribution in finite samples cannot be detected in a statistically
significant manner. Another possibility is to remove the discontinuity
with an asymmetry of the left and right tail distribution. The later
case is regularly observed in equity indices where up moves are more
likely but the negative returns have a larger tail (i.e. distribution
skewness).

\begin{figure}[p]
\begin{centering}
\includegraphics[width=0.5\textwidth]{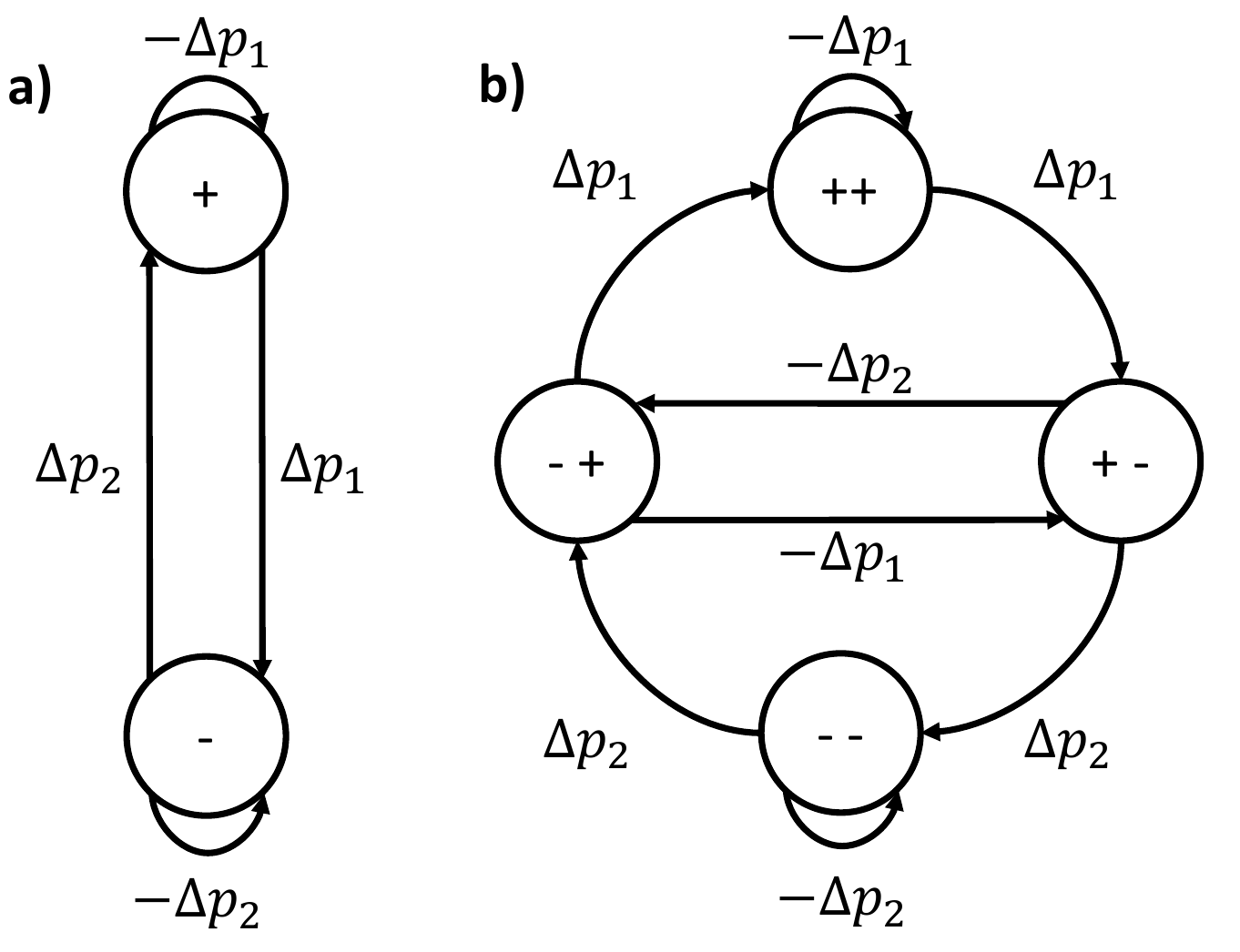}
\par\end{centering}
\caption{\textbf{Illustration of binary Markov processes and associated parameters:
a) one lag ($\protect\lag=1$); b) two lags ($\protect\lag=2$) and
stationary regime with equal probabilities for all states.\label{fig:Binary-Markov-process} }}
\end{figure}

Figure \ref{fig:Binary-Markov-process}.a shows the binary Markov
process with lag one ($\lag=1$). Figure \ref{fig:Binary-Markov-process}.b
shows the two degrees of freedom in the binary Markov process with
lag two ($\lag=2$), uniform state probability, and equal number of
up and down moves. When picking $\Deltaprob[1]=0.5$ and $-0.5<\Deltaprob[2]\ll0$
the binary Markov process with two lags produces almost exactly the
XOR pattern. The binary Markov processes with three and four lags
have more then two degrees of freedom when fulfilling the uniform
state probability and balancing conditions.

\subsection{Expected Autoregressive Forecast for Binary Markov Processes\label{subsec:expected-ar-markov}}

The Markov DGP defined in Equation (\ref{eq:Markov-DGP}) violates
the Martingale condition 
\begin{equation}
E\left[\stochvar[t+1]|\stochvar[t-\lag+1],\,\ldots,\,\stochvar[t]\right]=0
\end{equation}
when there exists a state $\state$ such that $\Deltaprob\neq0$.
We want to determine to which degree the deterministic pattern can
be predicted using an $\AR$ autoregressive model as defined in Equation
(\ref{eq:arp}). Assuming homoskedasticity of the inputs $\slinputs$
(not to be confused with the samples $\samples$) and outputs $\sloutputs$,
the unbiased least square regression estimator reads
\begin{equation}
\hat{\phi}=\left(\slinputs^{T}\slinputs\right)^{-1}\slinputs^{T}\sloutputs.
\end{equation}
To compute the expectation of this estimator for the binary Markov
DGP we notice that for every state $S\in\states=\bincat^{\lag}$ there
is a reverse state $-S$. Therefore, the set of states can be split
into two as $\states=\states^{+}\boldsymbol{\cup}\states^{-}$, where
$\left|\states^{+}\right|=\left|\states^{-}\right|$ and $\state[][+]=-\state[][-],\,\forall1\leq i\leq2^{\lag-1}$.
Each state is paired to its reverse. Assuming $n$ observed samples,
the inputs $\slinputs$ belong to the space $\states^{\times n}$,
and the distribution of $\sloutputs$ is defined by Equation (\ref{eq:Markov-DGP}).
Using these assumptions we compute
\begin{alignat}{2}
E\left[\slinputs^{T}\sloutputs\right] & = & \sum_{i=1}^{2^{\lag-1}}\alpha\left(\pi_{i}^{+}\state[][+]p_{i+}^{+}+\pi_{i}^{-}\state[][-]p_{i-}^{-}\right),
\end{alignat}
where $\alpha$ is a normalization factor stemming from the half-normal
distribution of the variables. In the case of uniform probabilities
for all states, the expectation simplifies to
\begin{equation}
E\left[\slinputs^{T}\sloutputs\right]=\frac{\alpha}{2^{\lag}}\sum_{i=1}^{2^{\lag-1}}\state[][+]\left(p_{i+}^{+}-p_{i-}^{-}\right).\label{eq:ar-unpredictable}
\end{equation}
The condition $\Delta p_{i}^{+}=\Delta p_{i}^{-}$ implies $E\left[\slinputs^{T}\sloutputs\right]=0$
as $p_{i+}^{+}-p_{i-}^{-}=\frac{1}{2}\left(1+\Delta p_{i}^{+}-1-\Delta p_{i}^{-}\right)=0$,
and at the same time fulfills the balancing condition of Equation
(\ref{eq:balancing-equation}). As a result, the estimated parameters
$\hat{\phi}$ are zero, and the autoregressive model is unable to
predict the deterministic pattern in the Markov process.

The binary Markov process with lag $\lag=1$ only has the two states
$\bincat$, which can be split into $\states^{+}=\left\{ +\right\} $
and $\states^{-}=\left\{ -\right\} $, and is determined by the two
parameters $\Delta p_{1}^{+}$ ($=\Deltaprob[1]$) and $\Delta p_{1}^{-}$
($=\Deltaprob[2]$). This process has zero expectation for an autoregressive
model when $\Delta p_{1}^{+}=\Delta p_{2}^{+}$, which implies the
stationary regime $\stationary[+]=\stationary[-]=\frac{1}{2}$. However,
the balancing condition of Equation (\ref{eq:balancing-equation})
is incompatible as it is satisfied when $\Delta p_{1}^{+}=-\Delta p_{2}^{+}$.
Nonetheless, as discussed in Section \ref{subsec:Binary-Markov-DGP},
the balancing condition is not necessary as the return distribution
can be skewed.

The binary Markov process with two lags can satisfy the balancing
condition and have zero expectation for an autoregressive process.
The DGP including the degrees of freedom fulfilling these conditions
are shown in Figure \ref{fig:Binary-Markov-process}.b.

\subsection{Expected Fixed Tree Forecast for Autoregressive Processes\label{subsec:tree-ar-expected}}

Assuming an autoregressive DGP $\AR$, we want to determine how well
a fixed regression tree with $\lag$ lags is able to predict the deterministic
component arising when $\left\Vert \phi\right\Vert >0$. The fixed
regression tree assigns to every of the $2^{\lag}$ input states $\state=\left(\state[i,\,1],\,\ldots,\,S_{i,\,\lag}\right)\in\states$
the mean
\begin{alignat}{1}
\hat{c}_{i}=E\left[\stochvar[t+1]|\textrm{sign}\left(\stochvar[t-\lag+1]\right)=\state[i,\,1],\,\ldots,\,\textrm{sign}\left(\stochvar[t]\right)=S_{i,\,\lag}\right].
\end{alignat}
In an autoregressive process with normally distributed innovations,
the lagged variables are distributed as $x_{t}=\left(\stochvar[t-\lag+1],\,\ldots,\,\stochvar[t]\right)\thicksim\mathcal{N}\left(0,\,\Lambda\left(\arparam\right)\right)$,
where $\Lambda\left(\phi\right)\in\mathbb{R}^{\lag}$ is the covariance
matrix of the $\lag$ lags, assuming the independent innovations are
distributed as $a_{t}\thicksim\mathcal{N}\left(0,\,\sigma\right)$.

Due to the autocorrelation, there is a directional bias of the return
dependent of the previous $\lag$ returns. This bias can be computed
for a region $\region[i]=\left\{ x\in\mathbb{R}^{\lag}|x_{j}\state[i\,j]\geq0,\,\forall j\right\} $
to be
\begin{equation}
P\left(\sloutput\geq0|\slinput\in\region[i]\right)=\int_{\region[i]}\mathcal{N}\left(x|\,0,\,\Lambda\left(\arparam\right)\right)\int_{-\arparam\cdot x}^{+\infty}\mathcal{N}\left(a|\,0,\,\sigma\right)dxda.\label{eq:region-proba-bias}
\end{equation}
In the general case of an arbitrary autoregressive parameter $\arparam$,
this integral cannot be evaluated in closed form.

The simplest case $\AR[1]$ has the analytically solution
\begin{equation}
P\left(\sloutput\geq0|\slinput\in\region[+]\right)=\frac{1}{2}+\frac{1}{\pi}\arctan\left(\frac{\phi}{1-\phi^{2}}\right),
\end{equation}
where $\region[+]=\mathbb{R}^{+}$. As expected, the bias is independent
of the variance of the innovations. The directional bias for the region
$\region[-]=\mathbb{R}^{-}$ is obtained by symmetry as $P\left(\sloutput\geq0|\slinput\in\region[-]\right)=1-P\left(\sloutput\geq0|\slinput\in\region[+]\right)$.
Assuming the calibration data is sufficient for a decision tree to
predict the bias correctly every time, the directional accuracy of
the tree will be $P\left(\sloutput\geq0|\slinput\in\region[+]\right)$.
An $\AR[1]$ predictor with correct parameter $\arparam$ always makes
the same prediction as a decision tree in this scenario, and therefore
the two predictors have identical directional accuracy. As an example,
for $\arparam=0.1$ the directional accuracy is $\approx0.532$.

Determining the directional accuracy for the $\AR[2]$ case leads
to integrals over multivariate distribution that cannot be solved
in closed form. Hence, only numerical solutions are possible and would
require a lengthy computation analyzing each region separately. Nonetheless,
let us remark what happens in the regions $\region[+-]$ and $\region[-+]$
for the parameter choice $\arparam=\left(\phi_{0}=0,\,\arparam[1],\,\arparam[2]=\arparam[1]\right)$.
The value of $\arparam\cdot x$ for $x=\left(x_{1},\,x_{2}\right)\in\region[+-]$
is anti-symmetric with respect to the axis $x_{2}=-x_{1}$. As a consequence
of this anti-symmetry, the integral defined in Equation (\ref{eq:region-proba-bias})
takes the constant value $P\left(\sloutput\geq0|\slinput\in\region[i]\right)=\frac{1}{2}$,
and this region has no predictability bias for a fixed tree. However,
the regions $\region[++]$ and $\region[--]$ exhibit a predictability
bias similar to the $\AR[1]$ case, and therefore the fixed tree predictor
will have roughly half the directional accuracy of the autoregressive
predictor in this specific case. Hence, fixed trees can be a sub-optimal
choice to predict autoregressive processes with two lags or more.
Nonetheless, they are more robust then autoregressive models, which
are entirely unable to predict a binary Markov process. Additionally,
the CART algorithm can generate a regression tree approximating arbitrarily
well the autoregressive process, assuming sufficient calibration data
is available. In reverse, the autoregressive model is always too rigid
to capture the binary Markov process.

\section{Statistical Test\label{sec:Statistical-Test}}

\subsection{Multiple Testing Methodology}

In this study, we aim to determine the significance level at which
a strategy outperforms a given benchmark. In particular in the context
of multiple testing with correlated strategies, as we want to test
the predictive power of the decision tree and autoregressive models
for different lags and calibration window lengths. This goal can be
achieved using the stepwise multiple testing method developed by \citet{romanowolf05}.
It acts on the observed data matrix $\left(\stochvar[\time,\,\strategies+1]\right)_{t,\,s}$
(i.e. returns), with $1\leq t\leq\time$ time steps of the $1\leq s\leq\strategies$
different strategies. The last column $\strategies+1$ is reserved
for the benchmark. The distribution of a chosen performance metric
is computed using bootstrapped realizations of $\stochvar[\time,\,\strategies+1]$.
Then a stepwise procedure rejects as many null hypothesis as possible,
at a given significance level $\alpha$, without violating the familywise
error rate.

\subsection{Test Statistic}

To determine the risk adjusted performance of the strategies, we chose
the Sharpe ratio as performance metric. In particular, we want to
determine if the Sharpe ratio of a strategy is significantly better
then the benchmark, and we therefore use as test statistic the Sharpe
ratio difference with respect to the benchmark. The following computation
of the test statistic is based on our interpretation and implementation
of the method developed by \citet{RePEc:zur:iewwpx:320}.

\subsubsection{Studentized Sharpe Ratio Test Statistic}

Assuming two return sequences $\left(\stochvar[1,\,1],\,\ldots,\,\stochvar[T,\,1]\right)$
and $\left(\stochvar[1,\,2],\,\ldots,\,\stochvar[T,\,2]\right)$ of
length $\time$, with estimated means $\left(\hat{\mu}_{1},\,\hat{\mu}_{2}\right)$
and variances $\left(\hat{\sigma}_{1}^{2},\,\hat{\sigma}_{2}^{2}\right)$,
the estimated test statistic reads
\begin{equation}
\hat{\Delta}\left(\hat{\mu}_{1},\,\hat{\mu}_{2},\,\hat{\sigma}_{1},\,\hat{\sigma}_{2}\right)=\hat{\textrm{Sh}}_{1}-\hat{\textrm{Sh}}_{2}=\frac{\hat{\mu}_{1}}{\hat{\sigma}_{1}}-\frac{\hat{\mu}_{2}}{\hat{\sigma}_{2}}.\label{eq:sharpe-test-statistic}
\end{equation}
This estimated test statistic has to be taken cautiously because it
can be biased in the case of correlated or heteroskedastic returns,
and has a significant variance in finite samples. To mitigate these
biases, it is better to use the studentized test statistic 
\begin{equation}
\hat{\Delta}_{S}=\frac{\hat{\Delta}}{s\left(\hat{\Delta}\right)},\label{eq:studentized-sr}
\end{equation}
corrected for the estimation error $s(\hat{\Delta})$. Considering
the vector of estimated moments $\hat{\nu}=\left(\hat{\mu}_{1},\,\hat{\mu}_{2},\,\hat{\sigma}_{1},\,\hat{\sigma}_{2}\right)$,
with covariance matrix $\Psi$, the estimation error is computed as
\begin{equation}
s\left(\hat{\Delta}\right)=\sqrt{\frac{\nabla_{\nu}\hat{\Delta}\left(\hat{\nu}\right)^{T}\Psi\nabla_{\nu}\hat{\Delta}\left(\hat{\nu}\right)}{T}},\label{eq:studentized-error}
\end{equation}
where $\nabla_{\nu}\hat{\Delta}\left(\hat{\nu}\right)$ is the gradient
with respect to the moments $\nu$ of the test statistic defined in
Equation (\ref{eq:sharpe-test-statistic}), and the covariance matrix
of the moments is given by $\Psi=\lim_{T\rightarrow+\infty}E\left[\hat{\nu}^{T}\hat{\nu}\right]$.

\subsubsection{Robust Covariance Estimation}

In finite samples, the covariance matrix $\Psi$ needed to estimate
the studentized test statistic defined in Equation (\ref{eq:studentized-sr})
may be subject to biases as well. A Heteroskedasticity and Auto-Correlation
(HAC) robust estimation of $\Psi$ is obtained with a kernel $k$
as
\begin{equation}
\hat{\Psi}=\frac{\time}{\time-4}\sum_{j=-\time+1}^{\time-1}k\left(\frac{j}{\bandwidth}\right)\hat{\Gamma}_{\time}\left(j\right),
\end{equation}
where
\begin{equation}
\hat{\Gamma}_{T}\left(j\right)=\begin{cases}
\frac{1}{\time}\sum_{t=j+1}^{\time}\hat{\nu}_{t}^{T}\hat{\nu}_{t-j} & j\geq0\\
\frac{1}{\time}\sum_{t=-j+1}^{\time}\hat{\nu}_{t+j}^{T}\hat{\nu}_{t} & j<0
\end{cases}.
\end{equation}
Within this paper, we use the Quadratic-Spectral (QS) kernel, for
which the optimal bandwidth $S_{T}^{*}$ can be computed using automatic
methods derived by \citet{andrews1991} and \citet{RePEc:oup:restud:v:61:y:1994:i:4:p:631-653.}.
The optimal bandwidth for the QS-kernel reads
\begin{equation}
\bandwidth[][*]\approx1.32\left(\bandalpha\time\right)^{2/5},
\end{equation}
where $\bandalpha$ is a constant dependent on the DGP (e.g. $\AR$,~$\ARMA$,~$\MA$).
When using real financial data, we first regress an appropriate model
on the data, and then compute the constant $\bandalpha$ based on
simulations with the regressed model.

\subsection{Null Bootstrap\label{subsec:Null-Bootstrap}}

The studentized test statistic $\hat{\Delta}_{S}$ is a performance
metric fairly robust with respect to the error $s(\hat{\Delta})$
induced by the intrinsic covariance structure of the data and the
finite sample size. However, it does not account for the variance
across different realizations of the unknown underlying probability
distribution generating the observed data. To compute a confidence
interval, the unknown probability distribution is typically assumed
to be stationary, and new realizations are generated using the circular
block bootstrap of \citet{politis1992circular}. This bootstrap method
samples, with replacement, blocks $\left(\stochvar[\time,\,\strategies+1]\right)_{t_{0}\leq t\leq t_{0}+b,\,s}$
of size $b$ from the observed returns, to generate bootstrapped observations
$X_{T,\,S+1}^{*}$.

Bootstrapping the data breaks the correlation structure between the
blocks of size $b$, which limits the estimation of the covariance
matrix to the blocks of size $b$ and could introduce major biases
in the estimated test statistic. Nonetheless, this issue can be overcome
by finding the optimal block size for a given DGP. The block size
selection is performed by computing the estimated test statistic at
confidence level $\alpha$ in a situation where the true test statistic
is known. For example, let us consider two independent realizations
of length $\time$ of an $\AR$ with given parameter $\arparam$,
the first being the strategy and the second the benchmark. The two
realizations have identical expected studentized Sharpe ratio. Therefore,
at significance level $\alpha$, the test statistic should reject
for a fraction $1-\alpha$ of the bootstraps of the realizations the
hypothesis that the strategy has superior test statistic then the
benchmark. The optimal block size $b\left(\alpha\right)$ is hence
a function of the significance level $\alpha$. The pseudo-code for
the block size selection can be found in \citet{RePEc:zur:iewwpx:320}.

Given $m=1,\,\ldots,\,M$ bootstrap resamples $X_{T,\,S+1}^{*,\,m}$
with the optimal block size, the centered studentized test statistic
of strategy $s$ for bootstrap $m$ is
\begin{equation}
\hat{\Delta}_{S,\,s}^{*,\,m}=\frac{\left|\hat{\Delta}_{s}^{*,\,m}-\hat{\Delta}_{s}\right|}{s\left(\hat{\Delta}_{s}^{*,\,m}\right)},
\end{equation}
where the test statistic is always computed with respect to the benchmark
in column $S+1$. The individual p-values are computed as
\begin{equation}
\pvalue_{s}^{*,\,m}=\frac{\left\{ \hat{\Delta}_{S,\,s}^{*,\,m}\geq\hat{\Delta}_{S,\,s}\right\} }{M+1},\label{eq:indivudal-pv}
\end{equation}
which is the fraction of centered test statistics that exceed the
original test statistic.

A drawback of bootstrapping the returns of the strategies and benchmark
is the stationarity assumption. For this assumption to hold, the predicted
stock market has to be in the same regime during the entire test period,
which is an unrealistic requirement for long test periods (>10 years).
The fundamental drivers of market regimes, such as monetary policies
that a reassessed several times a year, change too frequently for
markets to be stationary over long time periods. Bootstraps over the
past two decades would include realizations that do not contain the
dotcom bubble and financial crisis. As a consequence, the stationarity
condition defeats the purpose of computing p-values with respect to
realizations similar to the real market.

To avoid the issue of non-stationarity of the returns, we instead
bootstrap the signals (long or short) of the strategies and keep the
returns of the benchmark unchanged in every bootstrap. For each bootstrap,
the returns of the strategies are computed based on the benchmark
returns and the bootstrapped signals. In this setting, the bootstrapped
test statistic is computed as in Equation (\ref{eq:studentized-sr}),
and is not centered. The obtained p-values describe the probability
that the randomized strategies with same number of long and short
positions, and same cross-strategy correlation structure of the signals,
achieve the observed performance of the actual strategies.

\subsection{Adjusting P-values for Multiple Testing}

The individual test statistics computed in Equation (\ref{eq:indivudal-pv})
do not correct for the multiple testing of several strategies simultaneously.
To adjust for multiple testing, we use the stepdown procedure of \citet{Romano2016}.
In a first step, the individual strategies are ordered by increasing
p-value. The indices $\left\{ r_{1},\,\ldots,\,r_{S}\right\} $ denote
the permutation of $\left\{ 1,\,\ldots,\,S\right\} $ that fulfills
$\pvalue_{r_{1}}\leq\pvalue_{r_{2}}\leq\ldots\leq\pvalue_{r_{S}}$.
In a second step, the smallest p-value in the $m$th resample of the
$S-j$ worst strategies is denoted by
\[
\textrm{min}_{\pvalue,\,j}^{*,\,m}=\min\left\{ \pvalue_{r_{j}}^{*,\,m},\,\ldots,\,\pvalue_{r_{S}}^{*,\,m}\right\} .
\]
Introducing the p-value $\pvalue_{0}^{adj}=0$, the adjusted p-values
can be computed recursively as
\[
\pvalue_{j}^{adj}=\max\left\{ \frac{\#\left\{ \textrm{min}_{\pvalue,\,j}^{*,\,m}\leq\pvalue_{j}\right\} +1}{M+1},\,\pvalue_{j-1}^{adj}\right\} 
\]
for $j=1,\,\ldots,\,S$. The precision of the computed p-values depends
on the number of null resamples $M$. For practical purposes $M=1000$
is sufficient to estimate the p-values at three decimal places.

\section{Simulation Study\label{sec:Simulation-Study}}

\subsection{Competing Strategies\label{subsec:Competing-Strategies}}

\begin{table}[p]
\begin{centering}
\begin{tabular}{lll}
\noalign{\vskip0.1cm}
\textbf{Model} & \textbf{Description} & \multicolumn{1}{c}{}\tabularnewline[0.1cm]
\hline 
\noalign{\vskip0.1cm}
$\boldsymbol{AR}$ & \textbf{A}uto-\textbf{R}egressive & \tabularnewline[0.1cm]
\cline{1-2} 
\noalign{\vskip0.1cm}
$\boldsymbol{RT_{MSE}}$ & \multirow{1}{*}{\textbf{R}egression \textbf{T}ree with MSE loss} & \multirow{2}{*}{{\small{}$\begin{array}{c}
\lag\in\left\{ 1,\,2,\,3,\,4\right\} \\
\calibrationL\in\left\{ 10,\,20,\,\ldots,\,500\right\} 
\end{array}$}}\tabularnewline[0.1cm]
\noalign{\vskip0.1cm}
$\boldsymbol{CT_{Gini}}$ & \textbf{C}lassification \textbf{T}ree with Gini index loss & \tabularnewline[0.1cm]
\cline{1-2} 
\noalign{\vskip0.1cm}
$\boldsymbol{FRT}$ & \textbf{F}ixed \textbf{R}egression \textbf{T}ree with mean prediction & \multirow{2}{*}{}\tabularnewline[0.1cm]
\noalign{\vskip0.1cm}
$\boldsymbol{FCT}$ & \textbf{F}ixed \textbf{C}lassification \textbf{T}ree with majority
vote & \tabularnewline[0.1cm]
\end{tabular}
\par\end{centering}
\caption{\textbf{Overview of the strategies compared in this study}. A strategy
is defined by a model, the order parameter $\protect\lag$, and the
calibration window length $\protect\calibrationL$. A total of $5\,\textrm{models}\times$$4\,\textrm{lags}$$\times50\,\textrm{lengths}=$$1000\textrm{\,strategies}$
are tested.\label{tab:method}}
\end{table}

Section \ref{sec:methods} presented the tree based models, and Section
\ref{sec:Comparing-Forecasting-Power} compared their forecasting
power to autoregressive models. We now define how these models are
used as a trading strategy on a return sequence $\left\{ \stochvar[t]\right\} _{1}^{\time}$
of length $\time$, which can be actual stock market returns or returns
generated by an arbitrary DGP. At each time step, a model $\model$
is calibrated on the past $\calibrationL$ returns to forecast one
step ahead as $\tilde{X}_{t+1}=\model\left(\left\{ X_{t}\right\} _{t-L+1}^{t}\right)$,
where $\tilde{X}_{t+1}$ is the forecast of the model. For the purpose
of this study\footnote{We remark that in general continuous trading signals can be constructed
based on the value of a regression forecast or the class probabilities
of a classifier. However, within this paper we evaluate forecasting
models only based on binary signals, as more complex trading strategies
are out-of-scope for the purpose of studying the predictive power
of decision trees.}, the forecast $\tilde{X}_{t+1}$ is then mapped to a long or short
trading signal defined by $\signal_{t+1}=\textrm{sign}\left(\tilde{X}_{t+1}\right)$.
This produces a sequence of binary signals $\left\{ \signal_{t}\right\} _{L+1}^{T}\in\left\{ -1,\,1\right\} ^{\time-\calibrationL-1}$
that define the corresponding trading strategy on the returns $\left\{ \stochvar[t]\right\} _{1}^{\time}$.
Hence, a strategy is defined by a model, the number of lags $\lag$,
and the calibration window length $\calibrationL$. Table \ref{tab:method}
presents an overview of the models and the parameters of the associated
strategies.

The first $\calibrationL$ returns in the sequence $\left\{ \stochvar[t]\right\} _{1}^{\time}$
are part of the calibration window of size $\calibrationL$, and only
the subsequent returns are forecast one step ahead using a rolling
window of size $\calibrationL$. Nevertheless, for further convenience,
we define by $\time$ the number of forecast returns, implicitly assuming
that $\calibrationL$ initial returns have been cropped before the
first forecast.

\subsection{Observed Model Data}

Section \ref{sec:Statistical-Test} described how to compute robust
p-values for a set of strategies based on their data matrix $\stochvar[\time,\,\strategies+1]$.
We now define how to generate this data matrix for a set of strategies
trading on a return sequence $\left\{ \stochvar[t]\right\} _{1}^{\time}$
of length $\time$. As established in \ref{subsec:Competing-Strategies},
a strategy is generically described by its sequence of signals $\left\{ \signal_{t}\right\} _{1}^{T}\in\left\{ -1,\,1\right\} ^{\time}$
determining the long or short positions on the returns $\left\{ \stochvar[t]\right\} _{1}^{\time}$.
The entries of the observed data matrix are obtained as $\stochvar[t,\,s]=\left\{ \stochvar[t]\right\} _{1}^{\time}\circ\left\{ \signal_{t,\,s}\right\} _{1}^{\time}$,
where $\signal_{t,\,s}$ is the trading signal of strategy $s$ at
time $t$, and $\circ$ denotes the Hadamard product. We implicitly
assume that $\calibrationL$ initial data points have been cropped
for the first forecast, and denote by $\stochvar[\time,\,\strategies]$
the data matrix of the observed returns from the $\strategies$ strategies.
As benchmark in the column $\strategies+1$, we use the buy-and-hold
strategy of the returns. The p-values for a given data matrix are
estimated using $5000$ bootstrap iterations.

\subsection{Simulation Setup}

As finite sample realizations of a DGP have large variance, the p-value
for a strategy on a DGP has to be computed as an average over multiple
runs. To obtain a reasonably small confidence interval, a p-value
has to be averaged over 5000 runs with 500 bootstrap iterations for
the data matrix of each run. The p-value for a given strategy and
DGP depends on three parameters, namely the autocorrelation parameters
$\arparam$ or $\Delta p$ of the DGP (implicitly defining the lag
$\lag$), the number of forecast returns $\time$, and the calibration
window length $\calibrationL$. Given the large computational expense
of a single p-value, and the large parameter space, a small subset
of simulations have been chosen to study the impact of the different
parameters.

Following the theoretical analysis of Section \ref{subsec:Connecting-Fixed-Trees},
we use the two competing data generating processes: $DGP_{1}$, an
autoregressive process $\AR[2]$ with the two degrees of freedom $\phi=\left(\phi_{1},\,\phi_{2}\right)$;
and $DGP_{2}$, a binary Markov based process $\BMP[2]$ defined in
Equation (\ref{subsec:Binary-Markov-DGP}), with the two degrees of
freedom $\Delta p=\left(\Delta p_{1},\,\Delta p_{2}\right)$ as shown
in Figure \ref{fig:Binary-Markov-process}.b. Realizations of both
processes are generated with normally distributed innovations $\innovation\sim\mathcal{N}\left(0,\,\sigma=1\right)$.

First, we study the behavior of the statistical significance as a
function of the autocorrelation parameters, expecting significance
to increase with autocorrelation. We simulate two years of trading
($\time=500$ days) with a calibration length of $\calibrationL=50$.

Second, we verify that significance increases with the duration $\time$
as true skill becomes less likely to be a result of luck. In particular,
we want to determine the typical $\time$ needed to achieve a given
significance level. We pick a moderately explosive process with autocorrelation
$\arparam=\Delta p=\left(0.2,\,0.2\right)$ and a calibration length
of $\calibrationL=50$.

Third, we explore the impact of the calibration length $\calibrationL$
on the significance level. In finite samples, the estimation error
of small autocorrelations is large, and therefore the calibration
window length plays an important role. To illustrate this behavior,
we chose a small autocorrelation $\arparam=\Delta p=\left(0.1,\,0.1\right)$
and a sufficient number of time steps $\time=500$.

\subsection{Results}

\begin{figure}[p]
\begin{centering}
\includegraphics[width=1\textwidth]{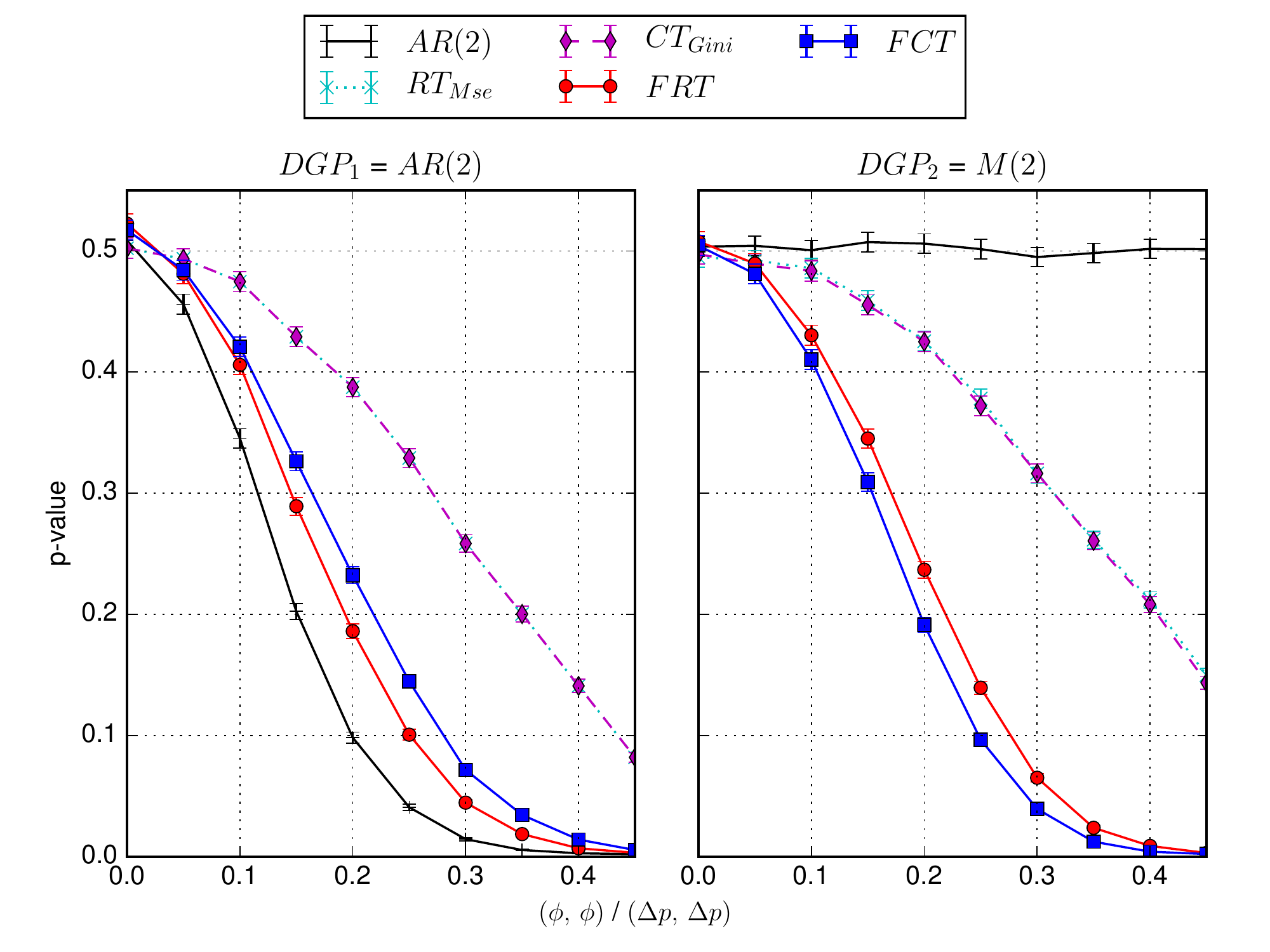}
\par\end{centering}
\caption{\textbf{P-values of all models when predicting an autoregressive process
$\protect\AR[2]$ and a binary Markov process $\protect\BMP[2]$ as
function of the parameters $\left(\protect\arparam[1],\,\protect\arparam[2]=\protect\arparam[1]\right)$
and $\left(\protect\Deltaprob[1],\,\protect\Deltaprob[2]=\protect\Deltaprob[1]\right)$.
}The prediction is made for sample size $\protect\time=500$ and calibration
window length $\protect\calibrationL=50$. The regression tree with
MSE loss and the classification tree with Gini index loss perform
similarly on both DGPs, but have the lowest performance on the autoregressive
processes. The fixed regression and classification tree perform almost
identically on both DGPs. The autoregressive process is optimal at
predicting itself, however fails entirely at predicting the binary
Markov process. The fixed trees exhibit a small bias at $\protect\arparam=0$
resulting from the effect described in Appendix \ref{sec:P-values-bias}.\label{fig:p-value-phi}}
\end{figure}

The simulated p-values as a function of the autocorrelation parameters
$\arparam$ and $\Delta p$ are presented in Figure \ref{fig:p-value-phi},
showing the increase in statistical significance as the autocorrelation
becomes stronger. To obtain a 95\% confidence level for an individual
strategy on a two year time frame, autocorrelations larger then $\left(0.3,\,0.3\right)$
need to be present.

\begin{figure}[p]
\begin{centering}
\includegraphics[width=1\textwidth]{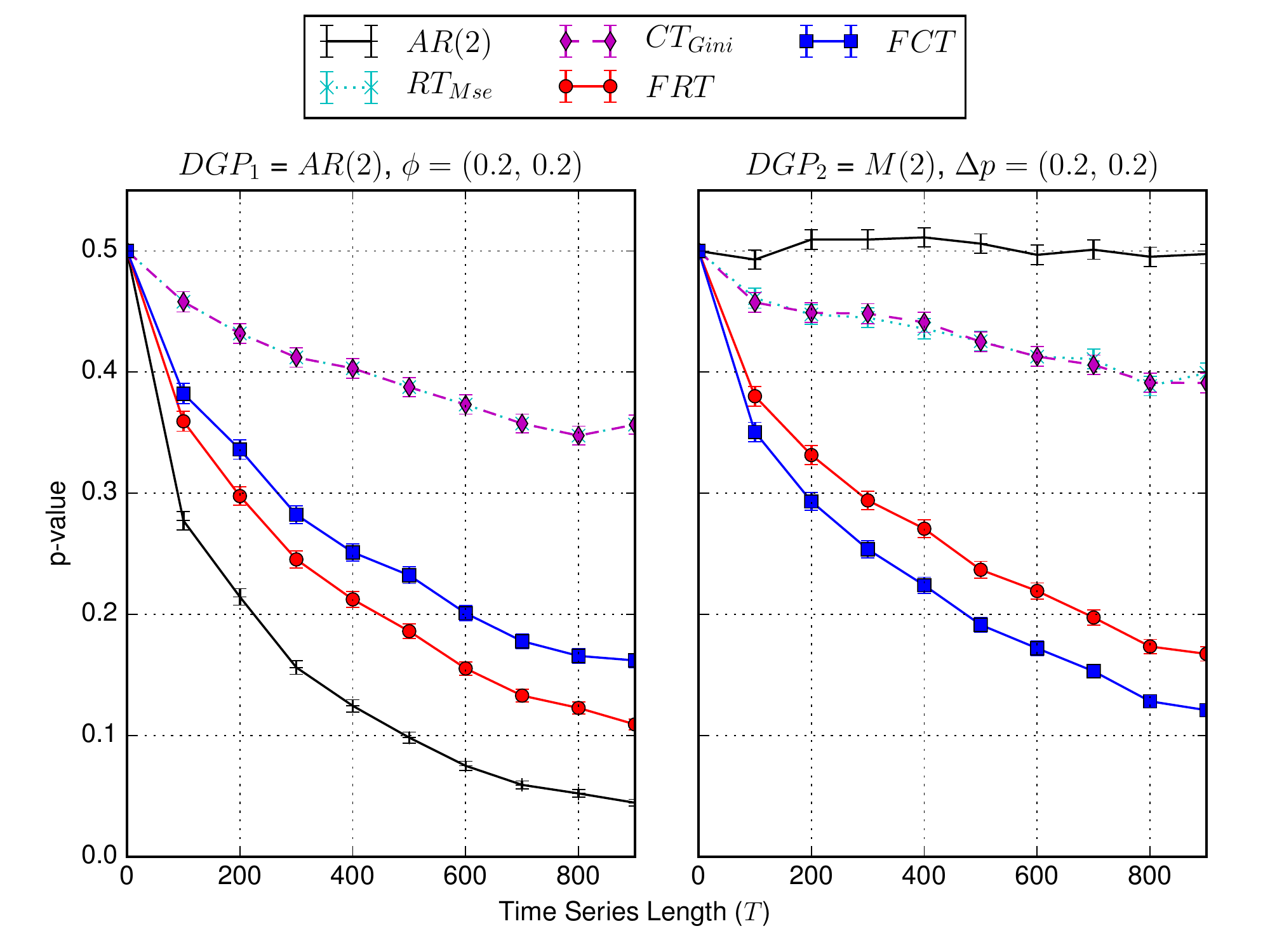}
\par\end{centering}
\caption{\textbf{P-values of all models with calibration length $\protect\calibrationL=50$
when predicting an autoregressive process $\protect\AR[2]$ and binary
Markov process $\protect\BMP[2]$, with parameters $\left(\phi_{1}=0.2,\,\phi_{2}=0.2\right)$,
respectively $\left(\protect\Deltaprob[1]=0.2,\,\protect\Deltaprob[2]=0.2\right)$.}
The significance of the performance increases with the sample size
$\protect\time$. The tree based models performs similarly on both
DGPs. The autoregressive predictor only work on the autoregressive
DGP.\label{fig:p-value-t}}
\end{figure}

Figure \ref{fig:p-value-t} shows the increase in statistical significance
with the sample size $\time$. Even for autocorrelation parameter
$\arparam=\left(0.2,\,0.2\right)$, which is much higher then average
on financial markets, at least $\time=800$ samples are needed before
a 95\% confidence level is achieved.

\begin{figure}[p]
\begin{centering}
\includegraphics[width=1\textwidth]{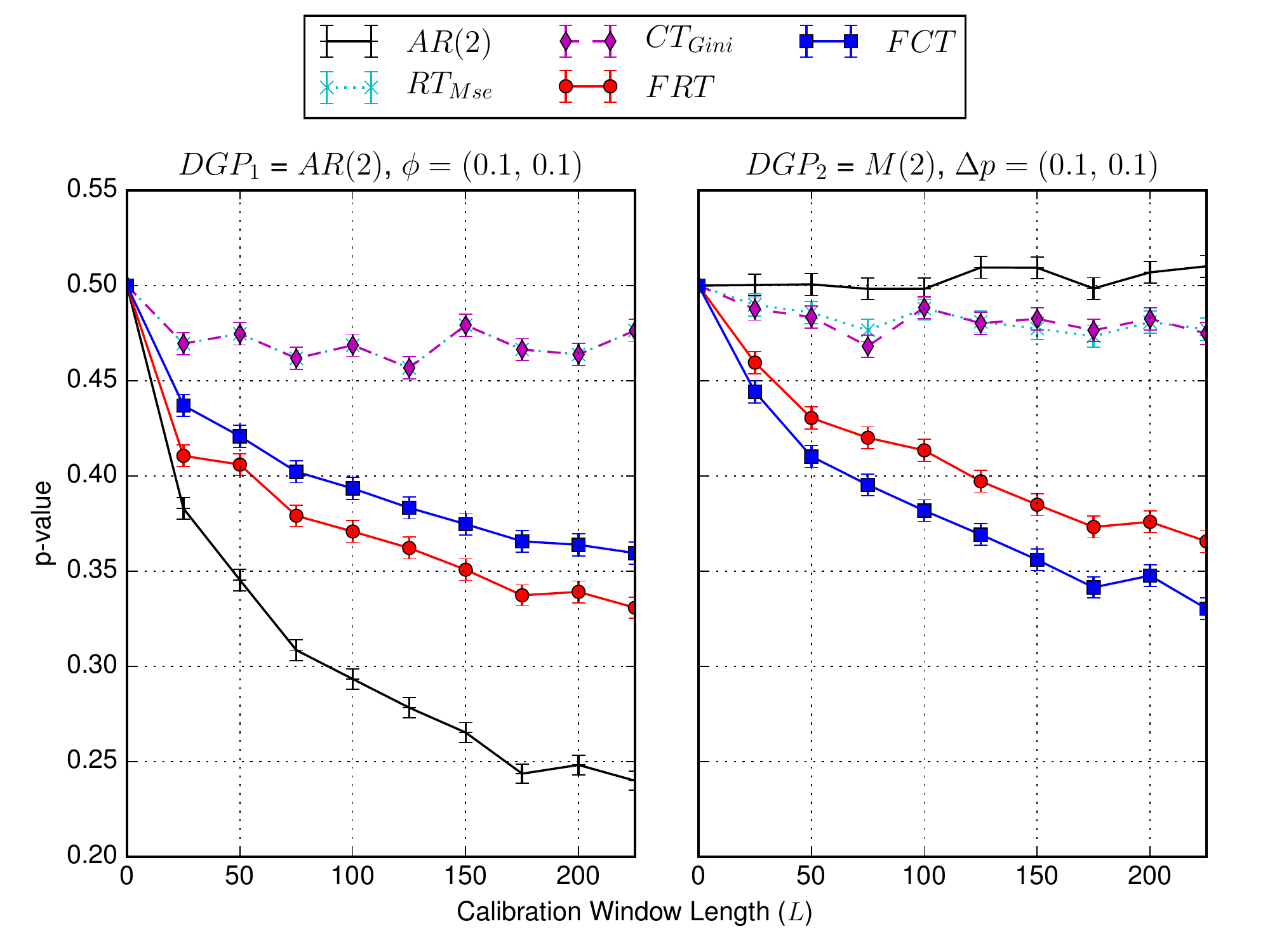}
\par\end{centering}
\caption{\textbf{P-values of all models for sample size $\protect\time=500$
when predicting an autoregressive process $\protect\AR[2]$ and binary
Markov process $\protect\BMP[2]$, with parameters $\left(\phi_{1}=0.1,\,\phi_{2}=0.1\right)$,
respectively $\left(\protect\Deltaprob[1]=0.1,\,\protect\Deltaprob[2]=0.1\right)$.}
The significance of the predictability increases with the calibration
window length $\protect\calibrationL$ as the estimation of the parameters
becomes more robust.\label{fig:p-value-l}}
\end{figure}

Figure \ref{fig:p-value-l} shows the decrease in significance with
the calibration window length $\calibrationL$. At least $\calibrationL=175$
samples for calibration are needed before the autocorrelation parameter
$\arparam=\left(0.1,\,0.1\right)$ can be estimated robustly.

As computed in Section \ref{subsec:expected-ar-markov}, the autoregressive
forecasting model does not pick up the deterministic pattern in the
binary Markov based process. The p-value converged to 0.5 for all
tested parameters. In contrast, the tree based performance is almost
identical on the autoregressive and binary Markov process. The regression
tree performs slightly better then the classification tree for the
autoregressive DGP, and vice versa for the binary Markov DGP. The
dynamic trees significantly underperform the fixed trees, especially
at small autocorrelations, as they overfit the noise.

\section{Empirical Results\label{sec:Empirical-Results}}

\subsection{Data \& Parameters}

To demonstrate how decision tree based models can outperform autoregressive
models on real data, we analyze the forecasting performance of all
strategies on daily returns of the S\&P 500 during the 20 year time
period Jan. 1, 1995 to Dec. 31, 2015. The daily returns are obtained
from Thomson-Reuters Eikon with dividend adjustment. The studied universe
of strategies arises from the models in Table \ref{tab:method}, backtested
for all calibration window lengths $\calibrationL\in\left\{ 10,\,20,\,\ldots,\,500\right\} $
and lags $\lag\in\left\{ 1,\,2,\,3,\,4\right\} $.

Two strategies that only differ by their calibration window lengths
$\calibrationL_{1}$ and $\calibrationL_{2}$ are highly correlated
when $\left|\calibrationL_{2}-\calibrationL_{1}\right|<10$. A step
size of $\Delta\calibrationL=10$ allows us to find the optimal length
with sufficient accuracy, while avoiding to compare almost identical
strategies. We remark that the bootstrap algorithm maintains the correlation
structure, and therefore the multiple testing adjusted p-values are
not impacted by the choice of $\Delta L$. The upper bound of 500
trading days on $\calibrationL$ results from the fact that the best
strategies were found well below this bound.

The limit $\lag\leq4$ on the number of lags $\lag$ arises from the
condition $\frac{\calibrationL}{2^{\lag}}\geq20$, which imposes the
lower bound of $20$ on the an average samples per leaf in a decision
tree, or equivalently a $\approx5\%$ accuracy on the class probabilities.
At lag $\lag=5$, the class probabilities would be determined with
a $6.5\%$ accuracy, which is too high for a meaningful prediction.

To perform the bandwidth selection for the HAC covariance matrix estimation,
as well as the bootstrap block size selection, we chose the parametric
$\AR-\textrm{GARCH}\left(\lag,\,\lag\right)$ model. The model parameters
are obtained by regression on the daily returns of the S\&P 500 on
the entire time period. The optimal parameters, determined by simulation,
were found to have a kernel bandwidth $S_{5000}^{*}=2.7$ and block
size $b=5$, roughly independent of the lag $\lag$.

\subsection{Performance of the Top Strategies}

The individual p-values of all 1000 strategies are shown in Figure
\ref{fig:p-value-overview}. The individual p-values are shown instead
of the multiple testing adjusted p-values because in the later most
p-values are simply one, and the relative performance is not visible
anymore. The best performing predictor is the fixed classification
tree ($FCT$) with lags $\lag=2$ and calibration length $L=370$,
which had a studentized Sharpe ratio performance of $\Delta_{S}^{1st}=2.12$,
noticeably higher then the second best predictor at $\Delta_{S}^{2nd}=2.05$.
This predictor was found to have a performance above the 99\% confidence
level when adjusting for multiple testing of all 1000 strategies.
The fixed classification and regression trees are the only predictors
that reach a performance above the 95\% confidence level for a large
range of lags and calibration lengths, when adjusted for multiple
testing. This thick set of outperforming rules shows the presence
of a robust signal, which has low probability of being a spurious
phenomenon due to data-snooping. At lag $\lag=4$, the length $L$
is to small for a robust calibration length and no predictor reaches
significant results. This confirms the choice of tested lags. The
results are in line with the simulation where the fixed trees were
the most robust predictors, and longer calibration length provide
a more robust parameter estimation.

The cumulative returns over time for the buy-and-hold strategy and
the best fixed classification tree predictors are shown in Figure
\ref{fig:top-cumulative-returns} without and with typical transaction
costs. The one-way transaction costs of 0.05\% applied at each buy
or sell is supported by the discussion of \citet[sec. 4.2]{Hsu2010}.
The strongest predictability arises during the burst of the dotcom
bubble, the crash of the financial crisis, and the European debt crisis.
Some performance metrics for the best strategy in each model class
are shown in Table \ref{tab:p-value-overview-table}. The best strategy
achieves twice the buy-and-hold return, twice the Sharpe ratio, and
roughly a third of the maximum draw down. The break even costs of
21.4 bps per round trip are higher then the typical 10 bps of costs
on actual markets .

\subsection{Further Analysis}

\begin{figure}[p]
\begin{centering}
\includegraphics[width=1\textwidth]{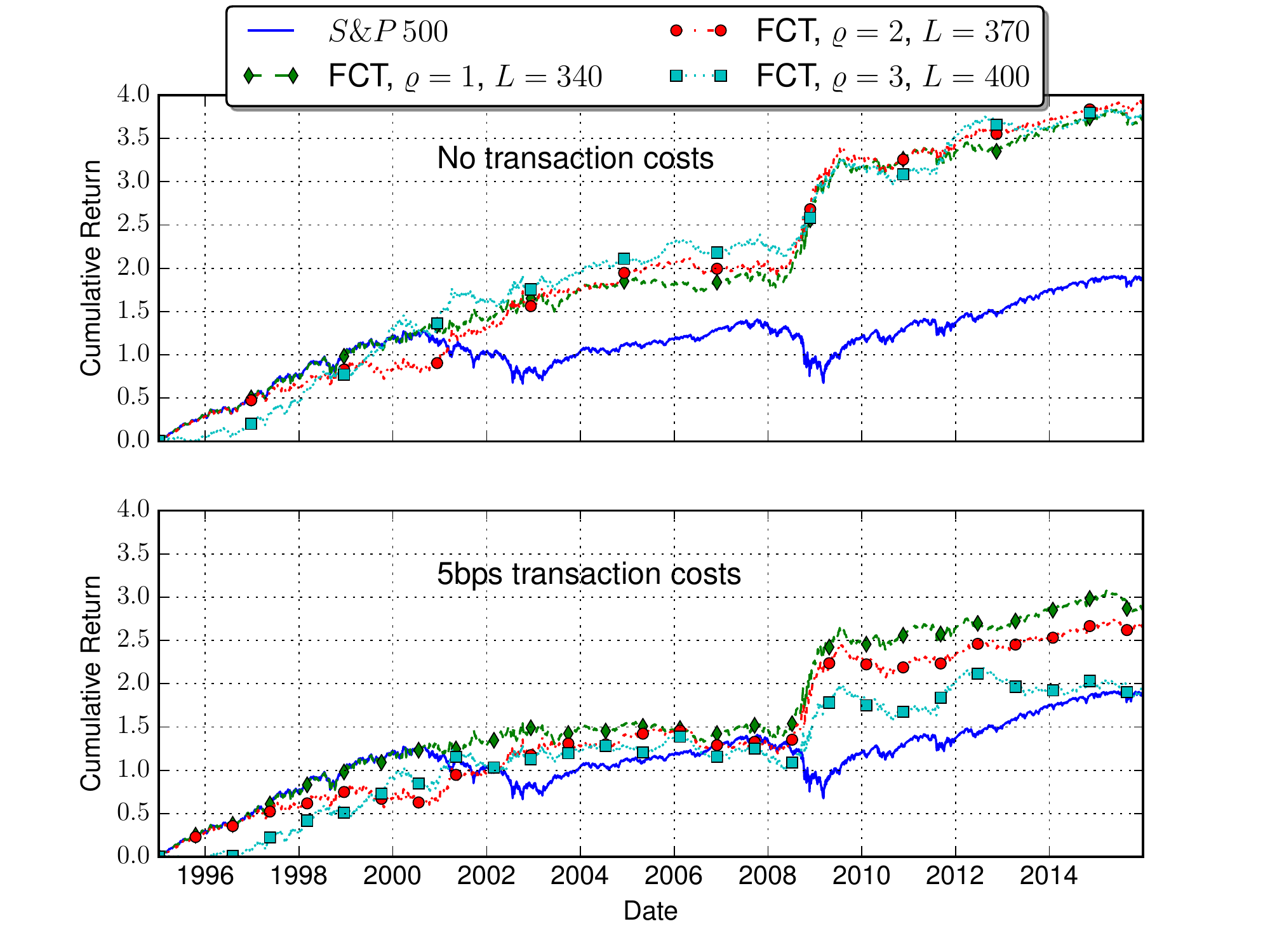}
\par\end{centering}
\caption{\textbf{Cumulative returns without (upper plot) and with (lower plot)
transaction costs of the top performing model ($FCT$) at three different
lags on the S\&P 500 during the 20 year time period Jan. 1, 1995 to
Dec. 31, 2015. }The highest predictability arises during the crash
of the financial crisis in 2008 for all lags. Noticeable predictability
arises as well during the implosion of the dotcom bubble from 2000
to 2003 at lag $p=2$ and during the European debt crisis in late
2011 at lag $p=3$. The lower plot shows the same strategies with
10bps of round trip transaction costs.\label{fig:top-cumulative-returns}}
\end{figure}

The statistical significance of the best performing strategies does
not imply that the EMH is violated, as it may have been impossible
to select one of these strategies ex-ante. Following the EMH definition
of \citet{Timmermann200415}, there needs to be a search technology
that would have selected the winning strategy. To test for the EMH,
we use the search technology that invests at every point in time into
the strategy with the best Sharpe ratio after 10bps of round trip
transaction costs. The search technology uses an expanding window
over all past returns. As can be seen in the lower plot of Figure
\ref{fig:top-cumulative-returns}, the FCT strategy with lag $\lag=1$
and $\calibrationL=340$ starts to outperform the two next best strategies
early on around 1997. The search technology confirms that no other
strategy in our universe of 1000 strategies would have hindered the
ex-ante generation of economic profits in excess of the buy-and-hold
strategy. Consequently, our result does seem to violate the EMH, or
at least raises questions about what limits to arbitrage could have
prevented the implementation of such strategies, at least in the last
two decades. The test period, which includes two major crashes, is
unlikely to hide extreme events that could drastically change the
downside risks of the strategy with respect to the market. We believe
that our test shows excess economic profits after realistic risk adjustments.
However, an accurate simulation of transaction costs and potential
market friction would have to be performed to confirm the result.

To determine if this abnormal performance can be explained by known
factors, we evaluate the performance with the CAPM, the three-factor
model of \citet{fama_french_1993}, and the four-factor model of \citet{Carhart1997}.
The full four-factor model measures performance as a time-series regression
of
\begin{equation}
\return-\riskfree[t]=\alpha+\betamarket\left(\rmarket[t]-\riskfree[t]\right)+\betasmb SMB_{t}+\betahml HML_{t}+\betamom MOM_{t}+e_{t}.
\end{equation}
In this regression, $\return$ is the strategy return on month $t$,
$\riskfree[t]$ is the risk-free rate (the 1-month U.S. Treasury bill
rate), $\rmarket[t]$ is the market return, $SMB_{t}$ and $HML_{t}$
are the size and value-growth returns of \citet{fama_french_1993},
$MOM_{t}$ is the momentum return, $\intercept$ is the average return
not explained by the benchmark model, and $e_{t}$ the residual error
term. The values for $\riskfree[t]$, $\rmarket[t]$, $SMB_{t}$,
$HML_{t}$ and $MOM_{t}$ are taken from Ken French's data library
\citep{french-data-library}, and derive from underlying stock returns
data from the Center for Research in Security Prices (CRSP). The three-factor
model is obtained by leaving out the momentum term, and the CAPM is
obtained by further leaving out the $SMB$ and $HML$ factors. Table
\ref{tab:Intercepts-and-slopes} shows the intercept and regression
slopes (load) for all three models, including their t-statistics.
The returns of the three best strategies are partially explained by
the market returns (significant $\betamarket$), but nevertheless
all three have significant intercept $\alpha$ that remains unexplained
by known factors. The return sign correlation uncovered by the decision
tree strategies qualifies as a new anomalous factor.

\begin{figure}[p]
\begin{centering}
\includegraphics[width=1\textwidth]{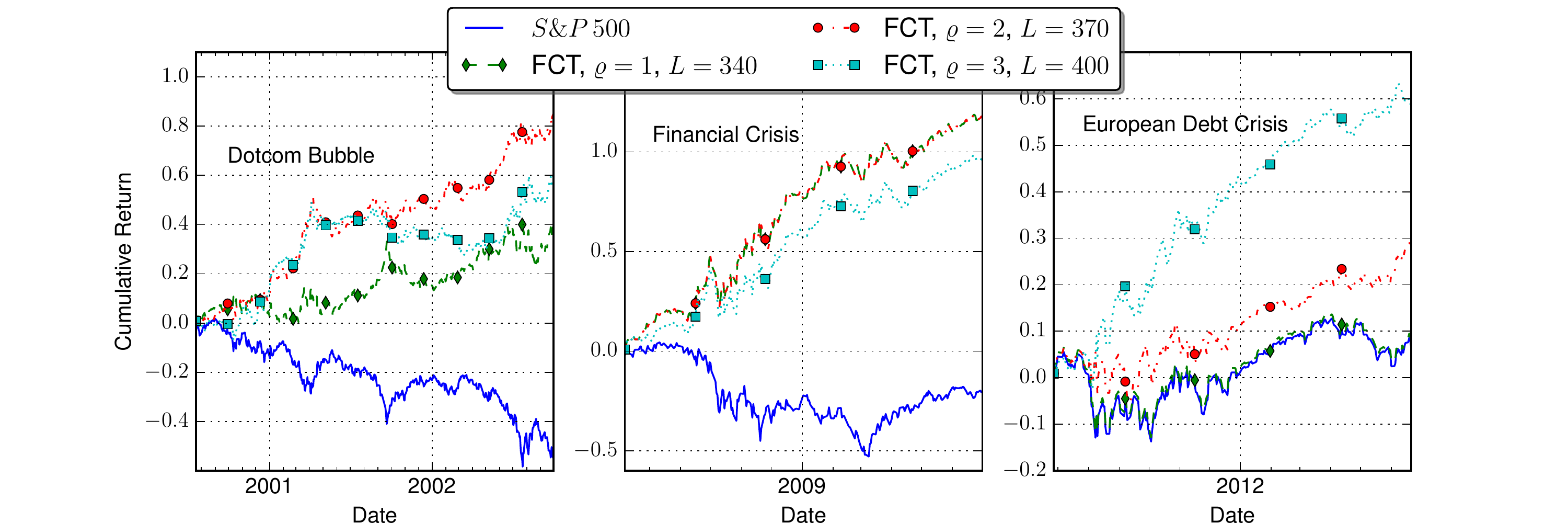}
\par\end{centering}
\caption{\textbf{Cumulative returns without transaction costs of the top performing
model ($FCT$) at three different lags on the S\&P 500 during the
anomalous periods.} The left plot shows the anomalous performance
of the fixed classification tree with lag $\protect\lag=2$ and calibration
length $\protect\calibrationL=340$ during the burst of the dotcom
bubble. The center plot shows the anomalous performance of the fixed
classification tree with lag $\protect\lag=1$ and calibration length
$\protect\calibrationL=370$ during the financial crisis. The right
plot shows the anomalous performance of the fixed classification tree
with lag $\protect\lag=3$ and calibration length $\protect\calibrationL=400$
during the European debt crisis.\label{fig:anomalies-zoom}}
\end{figure}

To obtain a better understanding of the return dynamics that generate
the predictability during the burst of the different bubbles, we present
in Figure \ref{fig:anomalies-zoom} a zoom-in on this periods. During
each period, the best performing strategy is significant at the 99.9\%
level when performing a bootstrap of returns under the stationarity
condition (see Section \ref{subsec:Null-Bootstrap}). The burst of
the dotcom bubble is dominated by a predictable return sign pattern
at lag two (break-even cost of 45 bps per round trip). The crash of
the financial crisis is dominated by a one day return sign dependency
(break-even cost of 83 bps per round trip). Finally, the European
debt crisis exhibits a more intricate three lag pattern that cannot
be captured by lower lag strategies (break-even cost of 32 bps per
round trip). The abnormal transition probabilities of these patterns
are presented in Table \ref{tab:Daily-return-sign}. These stationary
snapshots over the whole duration of each bubble confirm the significant
directional accuracy.

\begin{figure}[p]
\begin{centering}
\includegraphics[width=1\textwidth]{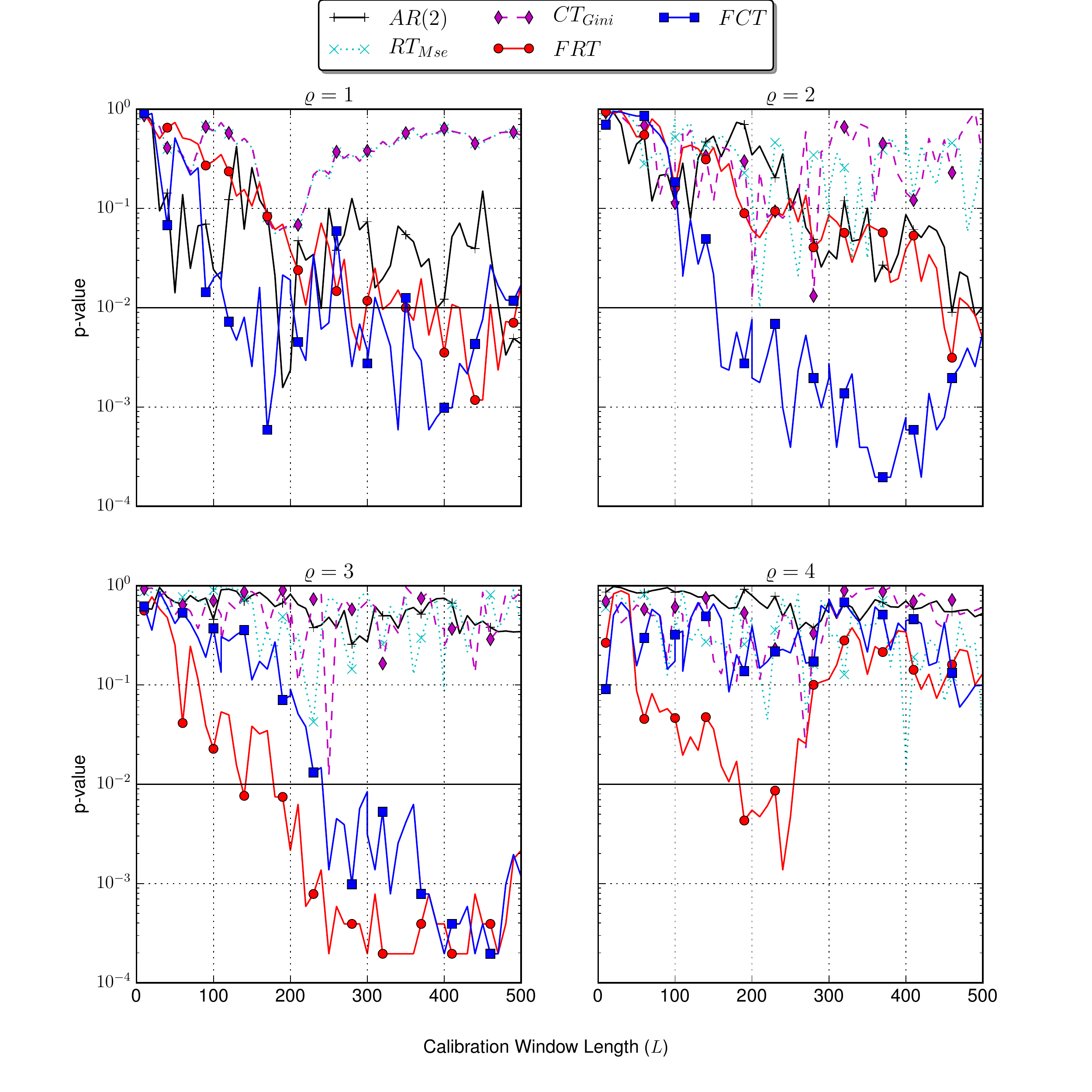}
\par\end{centering}
\caption{\textbf{Overview of the individual p-values of all strategies on the
S\&P 500 during the 20 year time period Jan. 1, 1995 to Dec. 31, 2015.
}The five forecasting models are: autoregressive ($AR$); regression
and classification tree ($RT_{MSE}$, $CT_{Gini}$); and the fixed
regression and classification trees ($FRT$, $FCT$). Each forecasting
model is run for the lags $\protect\lag=\left\{ 1,\,2,\,3,\,4\right\} $
and calibration window lengths $\protect\calibrationL=\left\{ 10,\,20,\,\ldots,\,500\right\} $.
The p-values are obtained by 5000 bootstrap simulations. The fixed
trees perform above the 95\% confidence level at lags $\left\{ 1,\,2,\,3\right\} $
and calibration window length $\protect\lag\geq250$. The other models
never perform significantly.\label{fig:p-value-overview}}
\end{figure}

Applying the theoretical model developed in Section \ref{subsec:Binary-Markov-DGP}
and \ref{subsec:expected-ar-markov}, to the stationary snapshots
of the three periods of abnormal returns, confirms the fundamental
concept that autoregressive models poorly capture return sign correlations.
The expected autoregressive parameters computed with Equation (\ref{eq:ar-unpredictable})
based on the values of Table \ref{tab:Daily-return-sign} are
\begin{align*}
E_{\textrm{Dotcom}}\left[\arparam\right] & =\left(0.006,\,-0.013\right),\\
E_{\textrm{Financial Crisis}}\left[\arparam\right] & =\left(-0.027\right),\,\textrm{and}\\
E_{\textrm{European Debt Crisis}}\left[\arparam\right] & =\left(-0.000,\,0.008,\,0.013\right).
\end{align*}
The expected parameters $\arparam$ have been scaled so as to be comparable
with the directional accuracies presented in Table \ref{tab:Daily-return-sign}.
The values show that the autoregressive model captures at most 2.7\%
of excess directional accuracy, far smaller then the maximal 17\%
of directional accuracy capture by the fixed classification tree.
Hence, the theoretical model explains why the autoregressive underperforms
the fixed trees on the S\&P 500 as shown by Figure \ref{fig:p-value-overview}.

\begin{table}[p]
\begin{centering}
\begin{tabular}{llllllllll}
\textbf{Model} & \textbf{$\lag$} & \textbf{$\calibrationL$} & $r_{y}$\textbf{\small{}(\%)} & $Sh_{y}$ & \textbf{MD (\%)} & $\pvalue$ & $\pvalue^{adj}$ & \textbf{BE }\textbf{\small{}(bps)} & \textbf{\#RT}\tabularnewline[0.1cm]
\hline 
\hline 
\noalign{\vskip0.05cm}
B\&H &  &  & 5.15 & 0.49 & -30.5 &  &  & 0 & 0\tabularnewline
\noalign{\vskip0.05cm}
FCT & 2 & 370 & 7.87 & 0.96 & -12.2 & $<2.0\cdot10^{-4}$ & $0.004$ & 16.4 & 1246\tabularnewline
\noalign{\vskip0.05cm}
FCT & 3 & 400 & 7.80 & 0.94 & -10.1 & $<2.0\cdot10^{-4}$ & $0.031$ & 10.6 & 1870\tabularnewline
\noalign{\vskip0.05cm}
FCT & 1 & 340 & 7.65 & 0.86 & -10.6 & $\leq5.9\cdot10^{-4}$ & $0.031$ & 21.4 & 860\tabularnewline
\noalign{\vskip0.05cm}
FRT & 3 & 330 & 7.86 & 0.92 & -12.6 & $<2.0\cdot10^{-4}$ & $0.043$ & 8.3 & 2417\tabularnewline
\noalign{\vskip0.05cm}
AR & 1 & 200 & 6.15 & 0.53 & -23.4 & $0.002$ & $0.750$ & 2.2 & 2728\tabularnewline
\noalign{\vskip0.05cm}
$RT_{MSE}$ & 2 & 210 & 5.75 & 0.55 & -13.7 & $0.010$ & $0.920$ & 1.3 & 2653\tabularnewline
\noalign{\vskip0.05cm}
$CT_{Gini}$ & 3 & 250 & 5.58 & 0.88 & -15.4 & $0.012$ & $0.953$ & 0.9 & 2651\tabularnewline
\end{tabular}
\par\end{centering}
\caption{\textbf{Summary of the top performing strategies.} The key values
in order are: the model family; the number of lags $\protect\lag$;
the calibration window length $\protect\calibrationL$; the compounded
annual return $r_{y}$; the yearly Sharpe ratio $Sh_{y}$; the maximum
draw down MD; the individual p-value; the p-value adjusted for multiple-testing
in the entire universe of models; the break-even transaction costs
(BE) with the buy-and-hold strategy; and the number of round trips
\#RT. \label{tab:p-value-overview-table}}
\end{table}

\begin{table}[p]
\begin{centering}
\begin{tabular*}{0.9\textwidth}{@{\extracolsep{\fill}}l|llllll}
\textbf{FCT} & $\intercept\,\left(t_{{\scriptscriptstyle \intercept}}\right)$ & ${\scriptstyle \betamarket\,\left(t_{MKT}\right)}$ & ${\scriptstyle \betasmb\,\left(t_{SMB}\right)}$ & ${\scriptstyle \betahml\,\left(t_{HML}\right)}$ & ${\scriptstyle \betamom\,\left(t_{MOM}\right)}$ & $R^{2}$\tabularnewline[0.2cm]
\hline 
\hline 
\noalign{\vskip0.1cm}
$\lag=1$ & 1.06 (3.5) & 0.31 (4.7) &  &  &  & 0.08\tabularnewline
$\calibrationL=340$ & 1.03 (3.4) & 0.31 (4.5) & 0.10 (1.1) & 0.10 (1.0) &  & 0.09\tabularnewline
 & 0.90 (2.9) & 0.38 (5.2) & 0.08 (0.8) & 0.16 (1.6) & 0.16 (2.7) & 0.11\tabularnewline
\hline 
$\lag=2$ & 1.22 (4.2) & 0.20 (3.1) &  &  &  & 0.04\tabularnewline
$\calibrationL=370$ & 1.20 (4.1) & 0.19 (2.9) & 0.07 (0.8) & 0.04 (0.4) &  & 0.04\tabularnewline
 & 1.21 (4.1) & 0.19 (2.6) & 0.08 (0.8) & 0.03 (0.3) & -0.02 (-0.3) & 0.04\tabularnewline
\hline 
$\lag=3$ & 1.19 (3.7) & 0.20 (2.7) &  &  &  & 0.03\tabularnewline
$\calibrationL=400$ & 1.20 (3.7) & 0.16 (2.2) & 0.13 (1.2) & -0.07 (-0.7) &  & 0.04\tabularnewline
 & 1.17 (3.5) & 0.18 (2.2) & 0.12 (1.2) & -0.06 (-0.6) & 0.04 (0.5) & 0.04\tabularnewline
\end{tabular*}
\par\end{centering}
\caption{\textbf{Intercepts and slopes in variants of regression for the three
top performing FCT strategies on the S\&P 500}. The table shows the
monthly intercepts ($\protect\intercept$) and regression slopes ($\protect\betamarket$,
$\protect\betasmb$, $\protect\betahml$ and $\protect\betamom$,
for $\protect\rmarket-r^{f}$, $SMB$, $HML$, and $MOM$, respectively),
as well as their t-statistics, for the CAPM, three-factor, and four-factor
versions of regression. The factors are estimated for the three top
performing FCTs on the S\&P 500 between Jan. 1, 1995 and Dec. 31 2015
at zero transaction cost, as shown in Figure \ref{fig:top-cumulative-returns}.
The monthly intercepts are significant for all strategies and regression
models. As well the market factor is significant in all cases, and
particularly strongly at lag $\protect\lag=1$. The FCT with lag one
correlates significantly with momentum.\label{tab:Intercepts-and-slopes}}
\end{table}

\begin{table}[p]
\begin{centering}
\begin{tabular}{ll|ll|ll}
\multicolumn{2}{l|}{Dotcom Bubble} & \multicolumn{2}{l|}{Financial Crisis} & \multicolumn{2}{l}{European Debt Crisis}\tabularnewline[0.1cm]
\hline 
\multicolumn{2}{l|}{$P\left(+\right)\approx0.48$} & \multicolumn{2}{l|}{$P\left(+\right)\approx0.52$} & \multicolumn{2}{l}{$P\left(+\right)\approx0.56$}\tabularnewline[0.1cm]
\hline 
$P\left(--\rightarrow+\right)$ & $\approx0.54$ & $P\left(-\rightarrow+\right)$ & $\approx0.61$ & $P\left(---\rightarrow+\right)$ & $\approx0.56$\tabularnewline[0.1cm]
$P\left(-+\rightarrow-\right)$ & $\approx0.53$ & $P\left(+\rightarrow-\right)$ & $\approx0.55$ & $P\left(--+\rightarrow+\right)$ & $\approx0.59$\tabularnewline[0.1cm]
$P\left(+-\rightarrow-\right)$ & $\approx0.54$ & \multirow{6}{*}{} &  & $P\left(-+-\rightarrow-\right)$ & $\approx0.50$\tabularnewline[0.1cm]
$P\left(++\rightarrow-\right)$ & $\approx0.56$ &  &  & $P\left(+--\rightarrow-\right)$ & $\approx0.56$\tabularnewline[0.1cm]
\multirow{4}{*}{} &  &  &  & $P\left(-++\rightarrow+\right)$ & $\approx0.67$\tabularnewline[0.1cm]
 &  &  &  & $P\left(+-+\rightarrow+\right)$ & $\approx0.60$\tabularnewline[0.1cm]
 &  &  &  & $P\left(++-\rightarrow+\right)$ & $\approx0.67$\tabularnewline[0.1cm]
 &  &  &  & $P\left(+++\rightarrow-\right)$ & $\approx0.55$\tabularnewline[0.1cm]
\end{tabular}
\par\end{centering}
\caption{\textbf{Daily return sign correlations of the S\&P 500 during the
burst of the dotcom bubble, financial crisis, and European debt crisis.}
The sign correlations are computed based the whole time periods shown
in Figure \textbf{\ref{fig:anomalies-zoom}}, and do not account for
potential non-stationarity. The burst of the dotcom bubble had more
down moves then up moves, with significant two day directional accuracy.
The crash of the financial crisis is impregnated by a daily reversal
of the return sign, with up moves being 11\% more likely then down
moves after a down move. The European debt crisis exhibits a more
intricate three day sign correlation, with an up move being 10 to
17\% more likely then a down move after two up moves and one down
move in arbitrary order.\label{tab:Daily-return-sign}}
\end{table}

\section{Conclusion}

The EMH is an assumption about financial market at the heart of many
regulatory decisions. This hypothesis has been verified to hold true
for a large range of regressive forecasting models, technical trading
rules and asset portfolios. However, recent developments in statistical
learning have not yet undergone a rigorous test.

In this paper, we presented a common non linearly separable pattern
that can arise, but cannot be forecast using autoregressive models.
Decision tree models possess arbitrary flexibility and are well suited
to capture these non linearly separable patterns. The issue of overfitting
can be addressed with an adequate lower bound on the number of samples
per leaf in a decision tree. We provide a connection between fixed
decision trees and Markov chains. The presented class of binary Markov
processes with a deterministic component are proven to be unpredictable
with autoregressive models. In contrast, the fixed classification
trees only marginally underperform an autoregressive forecast on an
autoregressive DGP for most parameter choices. A simulation study
confirmed the theoretical results and the robustness of fixed classification
and regression trees.

The models are tested on daily returns of the S\&P 500 for different
lags and calibration window lengths, giving rise to a universe of
1000 strategies. The multiple testing adjusted p-value of each strategy,
benchmarked against the buy-and-hold strategy, is computed using the
methodology of \citet{romanowolf05} and \citet{RePEc:zur:iewwpx:320}.
The fixed classification tree (FCT) at lags $\lag\in\left\{ 1,\,2,\,3\right\} $
and calibration window length $L\in\left[300,\,460\right]$ are the
best strategies, significant above the 95\% confidence level. This
confirms the simulation results where the fixed trees were robust
predictors for autoregressive and Markov based processes. The analysis
showed that the theoretical model holds true on the S\&P 500 to explain
the performance difference between decision trees and autoregressive
strategies.

Without transaction costs, the best fixed tree strategies more then
double the cumulative return and Sharpe ratio of the buy-and-hold
strategy. They break even with the buy-and-hold strategy at transaction
costs as high as 21 bps per round trip. A simple best Sharpe ratio
search technology could have selected ex-ante the best performing
strategy. The strategies all have significant intercept for the four-factor
regression model. Therefore, the EMH appears to be violated, in particular
during the dotcom bubble (2000-2003), the financial crisis (2008-2009)
and the European debt crisis (2012). During bull markets, the performance
of decision tree based strategies is roughly equal to the buy-and-hold
strategy. While no certain explanation can be given, it would not
be surprising to find that behavioral heuristics, as exposed by \citet{Tversky1974}
even among trained statisticians, have significant impact during a
market crash. The strong return sign correlations speak in favor of
biases in human heuristics relying more on past return signs then
on the amplitude.

The finding of this study stands in contrast with multiple prior market
efficiency studies that included the S\&P 500. The studies by \citet{Sullivan1999,citeulike:322540}
found no significantly performing technical trading rule on the S\&P
500, while our fixed classification trees perform significantly on
a longer test period of 20 years. The study by \citet{Hsu2010}, as
well testing technical trading rules, further concluded that market
efficiency has increased after the introduction of ETFs in the year
2000, and that emerging markets are less efficient then more mature
markets. Our study provides a solid counter example of a mature market
that exhibits large inefficiencies. As well, the largest inefficiencies
appear long after the introduction of ETFs, casting doubt on the impact
of ETFs on market efficiency. The discrepancy with prior studies is
a result of their focus on technical trading rules. Our work shows
that market efficiency cannot be measured only using technical trading
rules.

Noticeable research has gone into detecting explosive regimes in stock
markets \citep{Phillips2011,Kaizoji2011,SornetteCauwels2014}, and
studying the growth phase of bubbles. In contrast, the return dynamics
during the burst of a bubble seem under researched as shown by the
recent findings of an acceleration factor \citep{Ardila-Alvarez2015}
and the novel return sign predictability established in this paper.
Future research needs to study more in detail the predictability dynamics
during bubbles on other stock indices. As well, it needs to be better
understood which stocks in an equity drive the predictability.

\section*{Acknowledgments}

In particular, the authors would like to thank Z. Cui, D. Ardila,
Y. Malevergne, V. Filimonov, Q. Zhang, D. Daly and R. Kohrt for their
discussions, reviews and inputs.

\section*{Funding}

This work was supported by ETH Z\"urich under Grant {[}number 0-20029-14{]}.

\appendix

\section{P-value Bias\label{sec:P-values-bias}}

\begin{figure}[p]
\begin{centering}
\includegraphics[width=1\textwidth]{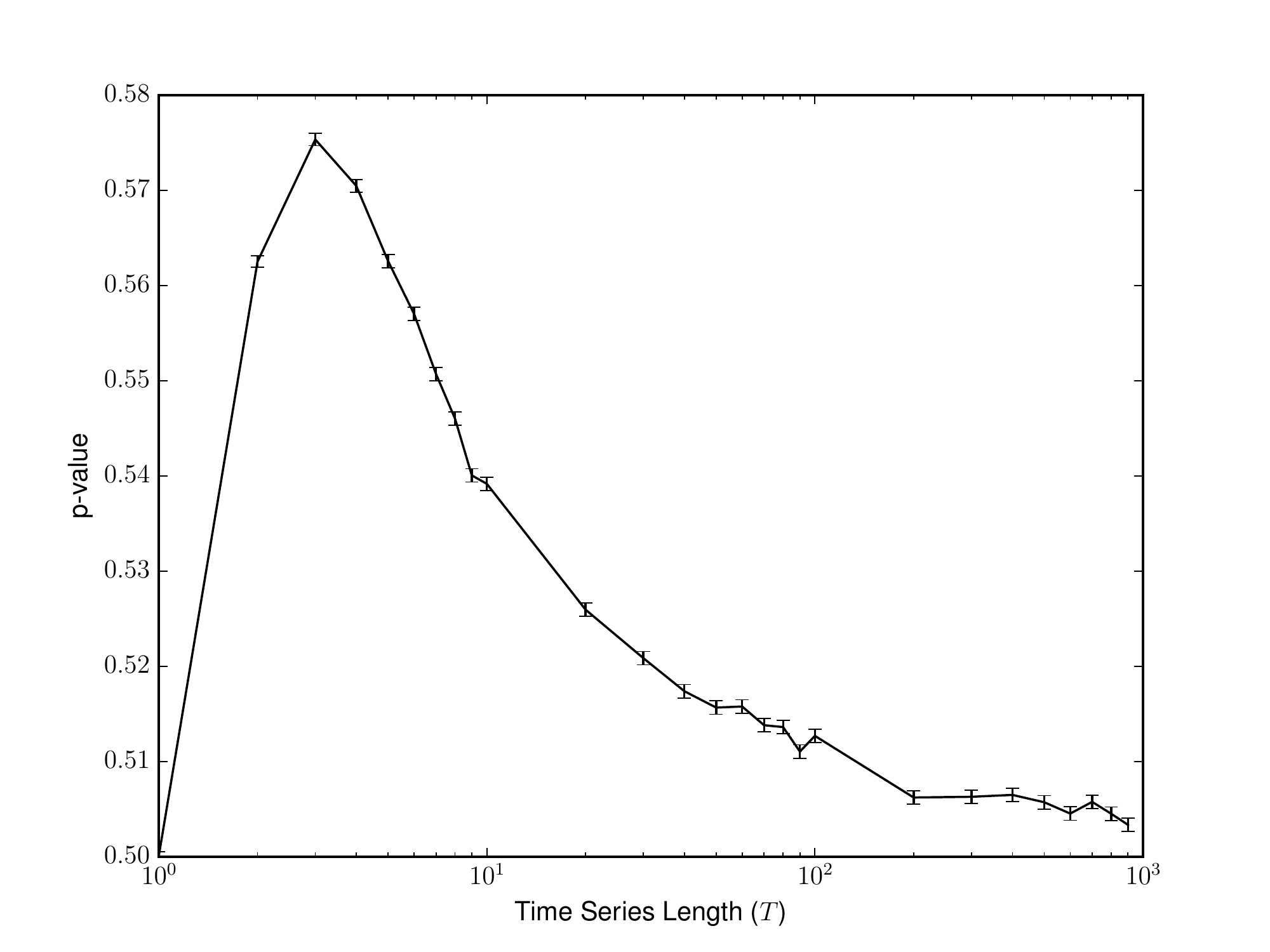}
\par\end{centering}
\caption{\textbf{P-value of the mean return metric for the previous sign prediction
strategy as a function of the sample size $\protect\time$.} As computed
in Equation \ref{eq:p-value-asymmetry}, the p-value is exactly $\protect\pvalue=0.5625$
for $\protect\time=2$. The bias decays with the sample length, but
remains significant above 0.5\% at sample length $\protect\time$.\label{fig:brain-twister-plot}}
\end{figure}

The p-values arising for certain metrics from simple prediction strategies
can exhibit non-intuitive biases. For example, let us consider the
strategy with signal
\begin{equation}
s_{t+1}=\textrm{sign}\left(r_{t}\right),
\end{equation}
which predicts for each time step the sign of the previous return.
We benchmark this strategy against the buy-and-hold strategy $s_{t+1}^{BH}=1$
using the mean return as a performance metric. When predicting two
returns $\left(r_{0},\,r_{1},\,r_{2}\right)$, where $r_{0}$ is a
dummy return needed for the prediction, the mean return difference
between the two strategies is
\begin{equation}
\Delta\bar{r}=\underbrace{\frac{1}{2}\left(\textrm{sign}\left(r_{0}\right)r_{1}+\textrm{sign}\left(r_{1}\right)r_{2}\right)}_{\textrm{strategy}}-\underbrace{\frac{1}{2}\left(r_{1}+r_{2}\right)}_{\textrm{buy-and-hold}}.
\end{equation}
While the expected mean return difference is zero, $E\left[\Delta\bar{r}\right]=0$,
the probabilities of it being positive or negative are not symmetric.
To show that $p\left(\Delta\bar{r}>0\right)\neq p\left(\Delta\bar{r}<0\right),$
we compute the value for all possible scenarios in Table \ref{tab:Mean-return-difference},
and can determine the probabilities to almost surely be
\begin{equation}
p\left(\Delta\bar{r}>0\right)=\frac{5}{16},\,p\left(\Delta\bar{r}=0\right)=\frac{4}{16}\;\textrm{and}\;p\left(\Delta\bar{r}<0\right)=\frac{7}{16}.\label{eq:p-value-asymmetry}
\end{equation}
The p-value associated to the strategy outperforming the benchmark
would be $p\left(\Delta\bar{r}<0\right)+\frac{1}{2}p\left(\Delta\bar{r}=0\right)=\frac{7}{16}+\frac{1}{2}\cdot\frac{4}{16}=0.5625$.
This bias decreases with the number of predicted returns as shown
in Figure \ref{fig:brain-twister-plot}. The bias decays with the
sample size $T$, but remains significant for the sample sizes of
interest.

\begin{table}[p]
\begin{centering}
\begin{tabular}{ll}
\toprule 
$\textrm{sign}\left(r_{0},\,r_{1},\,r_{2}\right)$ & $\Delta\bar{r}$\tabularnewline
\midrule
\midrule 
$\left(+,\,+,\,+\right)$ & $r_{1}+r_{2}-r_{1}-r_{2}=0$\tabularnewline
\midrule 
$\left(+,\,+,\,-\right)$ & $r_{1}+r_{2}-r_{1}-r_{2}=0$\tabularnewline
\midrule 
$\left(+,\,-,\,+\right)$ & $\frac{1}{2}\left(r_{1}-r_{2}-r_{1}-r_{2}\right)=\boldsymbol{-r_{2}\leq0}$\tabularnewline
\midrule 
$\left(+,\,-,\,-\right)$ & $\frac{1}{2}\left(r_{1}-r_{2}-r_{1}-r_{2}\right)=-r_{2}\geq0$\tabularnewline
\midrule 
$\left(-,\,-,\,-\right)$ & $\frac{1}{2}\left(-r_{1}-r_{2}-r_{1}-r_{2}\right)=-r_{1}-r_{2}\geq0$\tabularnewline
\midrule 
$\left(-,\,-,\,+\right)$ & $\frac{1}{2}\left(-r_{1}-r_{2}-r_{1}-r_{2}\right)=-r_{1}-r_{2}\sim0$\tabularnewline
\midrule 
$\left(-,\,+,\,-\right)$ & $\frac{1}{2}\left(-r_{1}+r_{2}-r_{1}-r_{2}\right)=\boldsymbol{-r_{1}\leq0}$\tabularnewline
\midrule 
$\left(-,\,+,\,+\right)$ & $\frac{1}{2}\left(-r_{1}+r_{2}-r_{1}-r_{2}\right)=\boldsymbol{-r_{1}\leq0}$\tabularnewline
\bottomrule
\end{tabular}
\par\end{centering}
\caption{\textbf{Mean return difference between the buy-and-hold and previous
sign prediction strategies}. The table lists all cases predicting
two returns $r_{1}$ and $r_{2}$.\label{tab:Mean-return-difference}}
\end{table}

\bibliographystyle{rQUF}
\bibliography{library}

\end{document}